\documentclass[a4paper,10pt]{revtex4}
\usepackage{graphicx}  % standard LaTeX graphics tool;
\usepackage{amsmath}   % For getting proper math eqns
\usepackage{amssymb}   % Used for getting various symbols eg. gtrsim, lesssim etc.
\usepackage{bm} % For bold math (esp of lower greek letters)
\usepackage{dcolumn}% Align table columns on decimal point
\usepackage{color}
\usepackage{mathrsfs}
\usepackage{amsfonts}
\usepackage{varioref}
\usepackage{mathrsfs}
\usepackage{graphicx}
\usepackage{latexsym}
\usepackage{amsmath}
\usepackage{amssymb}
\usepackage{textcomp}
\usepackage{amsbsy}
\usepackage{graphics}
\usepackage{epstopdf}
\usepackage{color}
\usepackage[caption=false]{subfig}

\RequirePackage[colorlinks,citecolor=blue,urlcolor=magenta,linkcolor=blue]{hyperref}
\input epsf

\allowdisplaybreaks[4]

\begin{document}

\tolerance=5000

\title{Bounce universe with finite-time singularity}

\author{Sergei~D.~Odintsov,$^{1,2}$\,\thanks{odintsov@ieec.uab.es}
Tanmoy~Paul$^{3}$\,\thanks{pul.tnmy9@gmail.com}} \affiliation{ $^{1)}$ ICREA, Passeig Luis Companys, 23, 08010 Barcelona, Spain\\
$^{2)}$ Institute of Space Sciences (ICE, CSIC) C. Can Magrans s/n, 08193 Barcelona, Spain\\
$^{3)}$ Department of Physics, Chandernagore College, Hooghly - 712 136, India }

\begin{abstract}

This work explains how the presence of a Type-IV singularity (a mild singularity) can influence the dynamics of a bouncing universe. In particular, 
we examine bounce cosmology that appears with a Type-IV singularity in the context of a ghost free Gauss-Bonnet theory of gravity. 
Depending on the time of occurrence of the Type-IV singularity, three different cases may arise -- 
when the singularity occurs before the bounce or after the bounce 
or at the instant of the bounce respectively. However in all of these cases, we find that in the case 
when the singularity ``globally'' affects the spacetime, 
the scalar power spectrum becomes red tilted and the tensor-to-scalar 
ratio is too large to be consistent with the observational data. 
Based on these findings, we investigate a different bouncing scenario which also appears with a Type-IV singularity, 
and the singularity affects the spacetime ``locally'' around the time when it occurs. As a result, and unlike to the previous 
scenario, the perturbation modes in the second bouncing scenario are likely to generate far away from the bounce in the deep contracting phase. 
This finally results to the simultaneous compatibility of the observable quantities 
with the Planck data, and ensures the viability of the bounce model where the Type-IV singularity has local effects on the spacetime around the time of 
the singularity.

\end{abstract}

%\pacs{}

\maketitle
\section{Introduction}

At present, we are living in a cosmological era where, on one hand, we have several cosmological data that includes the scalar spectral index, 
tensor-to-scalar ratio which describe the early stage of the universe, as well as the late time equation of state parameter, the Om(z) parameter 
in regard to the dark energy era of the universe. However, on other hand, the modern cosmology is still riddled with the question that 
whether the universe started its expansion from a Big-Bang singularity or from a non-singular like bouncing scenario. 
%Such an intriguing 
%question arises from the following simple consideration: if we think backwards in time, then the scale factor of the universe 
%reduces with time and finally leads to two different possibilities -- (1) in the first possibility, the scale factor finally hits to zero, where 
%the Kretschmann scalar diverges and yields a curvature singularity, while (2) in the second possibility, the scale factor attains a minimum 
%with a non-zero value (instead of going to zero) and yields a contracting stage of the universe, due to which, the evolution of the universe becomes 
%singular free. 
Inflation is one of the cosmological scenarios that successfully describes 
the early stage of the universe, in particular, it solves the horizon and flatness problems, and most importantly it 
predicts an almost scale invariant curvature perturbation power spectrum that is well consistent with the recent Planck data 
\cite{Guth:1980zm,Linde:2005ht,Langlois:2004de,Riotto:2002yw,Baumann:2009ds}. However, extrapolating 
backwards in time, the inflationary scenario results to an initial singularity of the universe, known as the Big-Bang singularity where, 
due to the geodesic incompleteness, the spacetime curvature diverges at the point of singularity. Bouncing cosmology is one of the 
alternatives of inflation, that can generate a scale invariant curvature power spectrum, and in addition, the bounce scenario leads to 
a singular free evolution of the universe
\cite{Brandenberger:2012zb,Brandenberger:2016vhg,Battefeld:2014uga,Novello:2008ra,Cai:2014bea,deHaro:2015wda,Lehners:2011kr,Lehners:2008vx,
Cai:2016hea,Li:2014era,Brizuela:2009nk,Cai:2013kja,Quintin:2014oea,Cai:2013vm,Raveendran:2017vfx,Raveendran:2018yyh,Raveendran:2018why,
Koehn:2015vvy,Koehn:2013upa,Martin:2001ue,Khoury:2001wf,
Buchbinder:2007ad,Brown:2004cs,Peter:2002cn,Gasperini:2003pb,Creminelli:2004jg,Lehners:2015mra,
Mielczarek:2010ga,Lehners:2013cka,Cai:2014xxa,Cai:2007qw,Cai:2012va,Avelino:2012ue,Barrow:2004ad,Haro:2015zda,Elizalde:2014uba,Banerjee:2020uil,
Odintsov:2020zct}. In the present work, we are interested in bounce cosmology that appears with certain features which we will explain.

Among various bounce models proposed so far, the matter bounce scenario (MBS) earned a lot of popularity, due to the fact that it produces a scale 
invariant primordial power spectrum and also leads to a matter dominated universe at the late expanding phase 
\cite{deHaro:2015wda,Quintin:2014oea,Raveendran:2017vfx,Brandenberger:2009yt,deHaro:2014kxa,Qiu:2010ch}. 
However the MBS is plagued with some 
problems, like -- (1) generally the tensor-to-scalar ratio in MBS becomes too large 
to be consistent with the observational data \cite{Akrami:2018odb}; (2) the evolution of the universe during the contracting stage becomes unstable due 
to the growth of anisotropies, that leads to the BKL instability in MBS \cite{new1}; (3) the scale factor describing MBS clearly depicts that the universe 
undergoes through a deceleration phase at the late time, which is not consistent with the dark energy observation 
\cite{Perlmutter:1996ds,Perlmutter:1998np,Riess:1998cb}. Here it deserves mentioning that 
such problems get well resolved by suitable modifications of matter bounce scenario, and the success of the bounce cosmology becomes 
quite illuminative. In particular, the authors of \cite{Nojiri:2019lqw,Elizalde:2020zcb} proposed an extended bounce scenario in Lagrange multiplier F(R) theory or in ghost 
free Gauss-Bonnet theory, where the first problem gets resolved, however the second and third problems persist. The article in \cite{Raveendran:2018yyh,
Paul:2022mup} 
proposed an ekpyrotic bounce scenario which is free from the BKL instability and predicts an almost scale invariant curvature power spectrum, 
but the models are unable in explaining the dark energy issue. Recently, a smooth unified scenario 
from an ekpyrotic bounce to the dark energy era has been proposed in \cite{Nojiri:2022xdo}, which stands to be a viable bounce in respect to the Planck data and 
concomitantly resolves the BKL instability.

Beside the Big-Bang or the initial singularity, there are other types of finite time singularities in cosmology, which are classified and studied for the 
first time in \cite{Nojiri:2005sx}. The most severe one among them is the Big-Rip or Type-I singularity, however 
there are three more such finite time singularities, 
like Type-II, III or Type-IV singularity. For studies on these types of singularity, see 
\cite{Barrow:2015ora,Nojiri:2015fra,Nojiri:2015wsa,Barrow:2004xh,Barrow:2004hk,Nojiri:2015fia,Odintsov:2015zza,Odintsov:2015ynk,Odintsov:2021yva,
deHaro:2012xj,Brevik:2021wzs,Barrow:2009df,Bouhmadi-Lopez:2006fwq,Yurov:2007tw}. 
In the case of Type-I, II or Type-III singularity, the effective energy density and/or 
the effective pressure of the universe diverge at the point of singularity. However on contrary, in the Type-IV singularity, the effective energy density 
and the effective pressure of the universe remain finite when the singularity occurs, and thus the Type-IV singularity is the most milder one 
among the finite time singularities. In particular, unlike to the Big-Rip singularity, the geodesic incompleteness does not appear in the 
Type-IV singularity, and as a result, the universe can smoothly pass through a Type-IV singularity (if any), and moreover 
the Type IV does not lead to catastrophic events to the observable 
quantities. Therefore it is possible that the universe faced a Type-IV singularity in the past during its evolution, through which, 
it smoothly passed without any geodesic incompleteness. However, the presence of 
a Type-IV singularity may has significant influence on the evolution of the universe as well as on the generation era of the 
primordial perturbation modes, as the singularity can globally affect the Hubble parameter. Therefore, in the realm of bouncing cosmology, the important 
question that immediately arises is following:
\begin{itemize}
 \item What are the possible effects of a Type-IV singularity on an otherwise non-singular bounce scenario ? 
 Is there any way to get a viable bounce scenario even in presence of a Type-IV singularity ?
\end{itemize}
We will address these questions in the present work. For the gravity theory, we will consider the well formulated $f(R,\mathcal{G})$ theory 
which turns out to be ghost free by the presence of the Lagrange multiplier in the gravitational action, 
as developed in \cite{Nojiri:2018ouv} ($R$ is the Ricci scalar and $\mathcal{G}$ is the Gauss-Bonnet term). The cosmology 
of $f(R,\mathcal{G})$ gravity from various perspectives have been discussed in 
\cite{Nojiri:2005vv,Li:2007jm,Carter:2005fu,Bamba:2021wyx,Odintsov:2020sqy,Odintsov:2020zkl,
Bamba:2020qdj,Cognola:2006eg,Nojiri:2020wmh}. The GB coupling function in the present context 
will be considered in such a way that it satisfies a constraint equation like 
$\ddot{h} = \dot{h}H$ (where $h(t)$ is the GB coupling function and $H(t)$ is the Hubble parameter of the universe). Owing to such condition on 
$h(t)$, the speed of the gravitational wave turns out to be unity, and the model becomes compatible with the event GW170817. In such scenario, our main 
aim will be to examine how the presence of a Type-IV singularity (a mild singularity) can influence the dynamics of a bouncing universe. Here we would 
like to mention that some of our authors have studied bounce cosmology with a Type-IV singularity in the context of F(R) gravity in \cite{Odintsov:2015ynk}, 
particularly the authors of \cite{Odintsov:2015ynk} showed that the presence of a Type-IV singularity destroys the viability of a bounce scenario with respect to the Planck data. 
However in the present work, we will consider a ghost free Gauss-Bonnet theory of gravity, and importantly, we will address how a bounce scenario 
that appears with a Type-IV singularity can be made viable with the recent observational data. These make the present work essentially different 
than \cite{Odintsov:2015ynk}. We will consider two different bouncing scenarios in the current work, 
namely -- (1) in the first scenario, the Type-IV singularity 
``globally'' affects the spacetime (where the term ``global'' means that although the Type-IV singularity occurs at a finite time $t = t_s$, it 
can show its effects even to the asymptotic evolution of the Hubble parameter at the distant past as well as at the distant future), 
and (2) in the second scenario, the Type-IV singularity affects the spacetime ``locally'' around the time when it occurs. These two 
scenarios are qualitatively different, in particular, in the former scenario where the Type-IV singularity globally affects the spacetime, the comoving 
Hubble radius asymptotically goes to zero at both sides of the bounce and thus the perturbation modes generate near the bounce; while in the second bounce 
scenario where the Type-IV singularity locally affects the spacetime, the comoving Hubble radius asymptotically diverges to infinity at the distant past and 
consequently the primordial perturbation modes generate far away from the bounce in the deep contracting phase. We will perform the scalar and tensor 
perturbation in these two scenarios, which are quite different compared to one another. The possible implications will be discussed 
in appropriate places.

The paper is organized as follows: in Sec.~[\ref{SecII}], we briefly discuss about the ghost free $f(R,\mathcal{G})$ gravity compatible with GW170817. 
Then in Sec.~[\ref{sec-global}] and in Sec.~[\ref{sec-local}], we will describe two different bounce scenarios as mentioned above, and 
consequently will examine the possible effects of Type-IV singularity on the bouncing dynamics. The paper ends with some conclusion 
in Sec.~[\ref{sec_conclusion}]. Finally we want to clarify the notations and conventions that we will adopt in the subsequent sections. 
We will work with natural units and the metric signature will be mostly positive i.e $\left(-,+,+,+\right)$. A suffix 'b' with a quantity will refer 
to the quantity at the instant of bounce, and $t=t_s$ will denote the time when the Type-IV singularity occurs. 
An overprime with some argument will indicate the derivative 
with respect to the argument, otherwise an overprime will represent $\frac{d}{d\eta}$. Moreover an overdot will denote the derivative 
with respect to the cosmic time.
.

\section{Brief about ghost-free $f(R,\mathcal{G})$ gravity compatible with the GW170817 event\label{SecII}}

Let us start by recalling the essential features of the
ghost free $f(R,\mathcal{G})$ gravity theory developed in \cite{Nojiri:2018ouv}. We consider
$f(R,\mathcal{G}) = \frac{R}{2\kappa^2} + f(\mathcal{G})$ with a Lagrange multiplier term in the 
gravitational action, where the presence of the Lagrange multiplier can eliminate the ghost modes from the theory. In particular, 
the action is given by \cite{Nojiri:2018ouv},
\begin{equation}
\label{FRGBg19} S=\int d^4x\sqrt{-g} \left(\frac{1}{2\kappa^2}R +
\lambda \left( \frac{1}{2} \partial_\mu \chi \partial^\mu \chi +
\frac{\mu^4}{2} \right)
 - \frac{1}{2} \partial_\mu \chi \partial^\mu \chi
- h\left( \chi \right) \mathcal{G} - V\left( \chi \right) +
\mathcal{L}_\mathrm{matter}\right)\, ,
\end{equation}
with $\mu$ being a constant having mass dimension $=[+1]$. By varying the action 
with respect to the Lagrange multiplier $\lambda$, one obtains a constraint equation as follows,
\begin{equation}
\label{FRGBg20} 0=\frac{1}{2} \partial_\mu \chi \partial^\mu \chi
+ \frac{\mu^4}{2} \, .
\end{equation}
The kinetic term of $\chi$ seems to be a constant and thus it can be captured in the scalar potential, as
\begin{equation}
\label{FRGBg21} \tilde V \left(\chi\right) \equiv \frac{1}{2}
\partial_\mu \chi \partial^\mu \chi + V \left( \chi \right)
= - \frac{\mu^4}{2} + V \left( \chi \right) \, ,
\end{equation}
and, as a result, the action of Eq.~(\ref{FRGBg19}) can be equivalently expressed as
\begin{equation}
\label{FRGBg22} S=\int d^4x\sqrt{-g} \left(\frac{1}{2\kappa^2}R +
\lambda \left( \frac{1}{2} \partial_\mu \chi \partial^\mu \chi +
\frac{\mu^4}{2} \right) - h\left( \chi \right) \mathcal{G}
 - \tilde V\left( \chi \right) + \mathcal{L}_\mathrm{matter}\right)
\, .
\end{equation}
The equations of motion for $\chi$ and $g^{\mu\nu}$ from the action (\ref{FRGBg22}) take the following forms,
\begin{align}
\label{FRGBg23} 0 =& - \frac{1}{\sqrt{-g}} \partial_\mu \left(
\lambda g^{\mu\nu}\sqrt{-g}
\partial_\nu \chi \right)
- h'\left( \chi \right) \mathcal{G} - {\tilde V}'\left( \chi \right) \, , \\
\label{FRGBg24} 0 =& \frac{1}{2\kappa^2}\left(- R_{\mu\nu} +
\frac{1}{2}g_{\mu\nu} R\right) + \frac{1}{2} T_{\mathrm{matter}\,
\mu\nu}
 - \frac{1}{2} \lambda \partial_\mu \chi \partial_\nu \chi
 - \frac{1}{2}g_{\mu\nu} \tilde V \left( \chi \right)
- D_{\mu\nu}^{\ \ \tau\eta} \nabla_\tau \nabla_\eta h \left( \chi
\right)\, ,
\end{align}
where $D_{\mu\nu}^{\ \ \tau\eta}$ has the form,
\begin{eqnarray}
 D_{\mu\nu}^{\ \ \tau\eta}&=&\big(\delta_{\mu}^{\ \tau}\delta_{\nu}^{\ \eta} + \delta_{\nu}^{\ \tau}\delta_{\mu}^{\ \eta} -
 2g_{\mu\nu}g^{\tau\eta}\big)R + \big(-4g^{\rho\tau}\delta_{\mu}^{\ \eta}\delta_{\nu}^{\ \sigma}
 - 4g^{\rho\tau}\delta_{\nu}^{\ \eta}\delta_{\mu}^{\ \sigma} + 4g_{\mu\nu}g^{\rho\tau}g^{\sigma\nu}\big)R_{\rho\sigma}\nonumber\\
 &+&4R_{\mu\nu}g^{\tau\eta} - 2R_{\rho\mu\sigma\nu} \big(g^{\rho\tau}g^{\sigma\nu} + g^{\rho\eta}g^{\sigma\tau}\big)
 \nonumber
\end{eqnarray}
with the expression $g^{\mu\nu}D_{\mu\nu}^{\ \ \tau\eta} =
4\bigg[-\frac{1}{2}g^{\tau\eta}R + R^{\tau\eta}\bigg]$. The trace of Eq.~(\ref{FRGBg24}) (i.e multiplying with $g^{\mu\nu}$) becomes
\begin{equation}
\label{FRGBg24A} 0 = \frac{R}{2\kappa^2} + \frac{1}{2}
T_\mathrm{matter} + \frac{\mu^4}{2} \lambda - 2 \tilde V \left(
\chi \right) + 4 \left( - R^{\tau\eta} + \frac{1}{2} g^{\tau\eta}
R \right) \nabla_\tau \nabla_\eta h \left( \chi \right) \, ,
\end{equation}
and solving  Eq.~(\ref{FRGBg24A}) with respect to $\lambda$ yields
\begin{equation}
\label{FRGBg24AB} \lambda = - \frac{2}{\mu^4} \left(
\frac{R}{2\kappa^2} + \frac{1}{2} T_\mathrm{matter}
 - 2 \tilde V \left( \chi \right) + 4 \left( - R^{\tau\eta}
+ \frac{1}{2} g^{\tau\eta} R \right) \nabla_\tau \nabla_\eta h
\left( \chi \right) \right) \, .
\end{equation}
The spatially
flat Friedmann-Robertson-Walker (FRW) metric ansatz will fulfill
our purpose in the present context, and hence we consider the line element as
\begin{equation}
\label{FRWmetric} ds^2 = - dt^2 + a(t)^2 \sum_{i=1,2,3} \left(
dx^i \right)^2 \, .
\end{equation}
Considering $\lambda$ and $\chi$ are homogeneous in cosmic time, i.e they depend on $t$ only, and also that $T_{\mathrm{matter}\, \mu\nu} =0$, 
then Eq.~(\ref{FRGBg20}) immediately results to the following solution
\begin{equation}
\label{frgdS4} \chi = \mu^2 \left(t-t_b\right) \, ,
\end{equation}
where $t_b$ is a constant and will be identified with the cosmic instance of bounce. 
Consequently, the temporal and spatial components of Eq.~(\ref{FRGBg24}) turns out to be,
\begin{align}
\label{FRGFRW1} 0 = & - \frac{3H^2}{2\kappa^2}
 - \frac{\mu^4 \lambda}{2} + \frac{1}{2} \tilde V \left( \mu^2 t \right)
 + 12 \mu^2 H^3 h' \left( \mu^2 t \right) \, , \\
\label{FRGFRW2} 0 = & \frac{1}{2\kappa^2} \left( 2 \dot H + 3 H^2
\right)
 - \frac{1}{2} \tilde V \left( \mu^2 t \right)
- 4 \mu^4 H^2 h'' \left( \mu^2 t \right) - 8 \mu^2 \left( \dot H +
H^2 \right) H h' \left( \mu^2 t \right) \, ,
\end{align}
and, furthermore, from Eq.~(\ref{FRGBg23}) we get
\begin{equation}
\label{FRGFRW3} 0 = \mu^2 \dot\lambda + 3 \mu^2 H \lambda - 24 H^2
\left( \dot H + H^2 \right) h'\left( \mu^2 t \right)
 - {\tilde V}'\left( \mu^2 t \right) \, .
\end{equation}
Here it may be mentioned that the above three equations are not independent, actually, by combining Eqs.~(\ref{FRGFRW4}) and
(\ref{FRGFRW3}), one can obtain Eq.~(\ref{FRGFRW2}). 
It is evident that Eq.~(\ref{FRGFRW1}) is an algebraic equation with respect to $\lambda$, on solving which, we get
\begin{equation}
\label{FRGFRW4} \lambda = - \frac{3 H^2}{\mu^4 \kappa^2} +
\frac{1}{\mu^4} \tilde V \left( \mu^2 t \right) + \frac{24}{\mu^2}
H^3 h' \left( \mu^2 t \right) \, .
\end{equation}
Similarly the scalar potential $\tilde V
\left( \mu^2 t \right)$ can be obtained by solving Eq.~(\ref{FRGFRW2}), and given by
\begin{equation}
\label{FRGFRW7} \tilde V \left( \mu^2 t \right) =
\frac{1}{\kappa^2} \left( 2 \dot H + 3 H^2 \right) - 8 \mu^4 H^2
h'' \left( \mu^2 t \right) - 16 \mu^2 \left( \dot H + H^2 \right)
H h' \left( \mu^2 t \right) \, .
\end{equation}
Recall that $\chi = \mu^2t$, and thus $\tilde V \left( \chi \right)$ being equal to
\begin{equation}
\label{FRGFRW8} \tilde V \left( \chi \right) = \left[
\frac{1}{\kappa^2} \left( 2 \dot H + 3 H^2 \right) - 8 \mu^4 H^2
h'' \left( \mu^2 t \right) - 16 \mu^2 \left( \dot H + H^2 \right)
H h' \left( \mu^2 t \right) \right]_{t-t_b=\frac{\chi}{\mu^2}}\, .
\end{equation}
The functional form of the Lagrange multiplier from Eq.~(\ref{FRGFRW4}) becomes
\begin{equation}
\label{FRGFRW4B} \lambda = \frac{2 \dot H}{\mu^4 \kappa^2} - 8 H^2
h'' \left( \mu^2 t \right) - \frac{8}{\mu^2} \left( 2 \dot H - H^2
\right) H h' \left( \mu^2 t \right) \, .
\end{equation}
Therefore a form of the Hubble parameter and the Gauss-Bonnet coupling function in turn fix the scalar potential and the Lagrange multiplier. 
We will consider the Hubble parameter in such a way that it leads to a bounce scenario with 
a Type-IV singularity. For the theory with
Lagrangian (\ref{FRGBg22}), it is well known that the speed of gravitational waves can be expressed as
\cite{Hwang:2005hb,Noh:2001ia,Hwang:2002fp},
\begin{eqnarray}
 c_T^2 = 1 + \frac{16\big(\ddot{h}-\dot{h}H\big)}{\frac{1}{\kappa^2} + 16\dot{h}H}
 \label{gravitaional wave speed}
\end{eqnarray}
with $H = \frac{\dot{a}}{a}$ being the Hubble parameter. Clearly the $c_T^2$ is different than unity, and the deviation of the $c_T^2$ from unity 
is controlled by the Gauss-Bonnet coupling function. However according to the GW170817 event (which validates the fact that the gravitational waves 
have same propagation speed as the electromagnetic waves which is unity in the natural units), we consider the Gauss-Bonnet coupling function in the 
present context in such a way that it results to $c_T^2 = 1$. For this purpose, we need to consider the coupling function by the following fashion 
\cite{Odintsov:2019clh},
\begin{eqnarray}
 \ddot{h} = \dot{h}H.
 \label{constraint on coupling}
\end{eqnarray}
Therefore to have a model compatible with GW170817, we will consider such Gauss-Bonnet coupling functions which satisfy Eq.~(\ref{constraint on coupling}). 
Finally, owing to the condition of 
Eq.~(\ref{constraint on coupling}), the scalar potential and the Lagrange multiplier can be simplified from Eq.~(\ref{FRGFRW8}) and 
Eq.~(\ref{FRGFRW4B}) respectively, and are given by,
\begin{eqnarray}
 \tilde{V}(\chi)&=&\left(2\dot{H} + 3H^2\right)\left(\frac{1}{\kappa^2} - 8\dot{h}H\right)~~,\label{final form1}\\
 \lambda(t)&=&\frac{2\dot{H}}{\mu^4}\left(\frac{1}{\kappa^2} + 8\dot{h}H\right)~~.
 \label{final form2}
\end{eqnarray}
Thus as a whole, Eq.~(\ref{constraint on coupling}), Eq.~(\ref{final form1}) and Eq.~(\ref{final form2}) 
are the main equations which, with a suitable form of $H(t)$, lead to the 
coresponding forms of $h(\chi)$, $V(\chi)$ and $\lambda(t)$ respectively. 

\section{Realization of a bounce with a Type-IV singularity}\label{sec-global}

As mentioned in the introductory section, we are interested to examine the possible effects of a mild singularity, particularly of a Type-IV singularity, 
in a bounce scenario. Therefore here 
we intend to realize a bounce scenario in presence of a Type-IV singularity, and depending on the time of singularity, three different cases 
may arise -- (1) when the singularity occurs before the bounce happens, (2) when the singularity occurs after the bounce and, (3) the case 
where the Type-IV singularity occurs at the instant of the bounce. The scale factor we consider, 
is given by,
\begin{eqnarray}
 a(t) = a_1(t)\times a_2(t) = \left(1+a_0\left(\frac{t}{t_0}\right)^2\right)^{n}\times 
 \exp{\left[\frac{f_0}{(\alpha+1)}\left(\frac{t-t_s}{t_0}\right)^{\alpha+1}\right]}~~.
 \label{scale factor-g}
\end{eqnarray}
The scale factor is taken as a product of two factors- $a_1(t)$ and $a_2(t)$ respectively, where $a_2(t)$ ensures the occurrence of a 
finite time singularity. 
Actually $a(t) = a_1(t)$ is sufficient for getting a non-singular bouncing universe where the bounce occurs at $t = 0$. However 
However for $a(t) = a_1(t)$, the bounce scenario becomes free from any finite time singularity. Thus due to our particular interest in the present 
context, i.e to examine the possible effects of a finite time singularity in an otherwise non-singular bouncing dynamics, 
we consider the scale factor as of Eq.(\ref{scale factor-g}) where $a_1(t)$ is multiplied by $a_2(t)$. The presence of $a_2(t)$ results 
to a finite time singularity 
at $t =t_s$ in the bouncing dynamics, as we will discuss after Eq.(\ref{der-H-g}). 
We will show that the presence of $a_2(t)$ does not harm the bouncing character of the universe, 
however it slightly shifts the bouncing time 
from $t = 0$ to a negative or a positive time depending on whether $t_s < 0$ or $t_s > 0$ respectively, 
and moreover the scale factor of Eq.(\ref{scale factor-g}) leads to an asymmetric bounce scenario (as $a(t) \neq a(-t)$). 
The parameter $\alpha$ present in the $a_2(t)$ is considered to have the form like $\alpha = \frac{2p+1}{2q+1}$ (with $p$ and $q$ are positive 
integers), so that the term $(t-t_s)^{\alpha+1}$ acquires positive values during the entire cosmic range 
(we take the root: $(-1)^{\alpha} = -1$ so that the scale factor and the corresponding Hubble parameter are real valued functions). 
Otherwise for $\alpha = \frac{2p}{2q+1}$, the 
term $(t-t_s)^{\alpha+1}$ becomes negative during $t<t_s$ and consequently $a(t) \rightarrow 0$ at the distant past, which is not healthy for a non-singular 
cosmological evolution of the universe. Thus we take $\alpha = \frac{2p+1}{2q+1}$ in the subsequent calculation. This is also important to get a bounce 
in the present context, as we will demonstrate below 
around Eq.(\ref{Hubble parameter-g-item1a}).

Eq.(\ref{scale factor-g}) immediately leads to the Hubble parameter and its first derivative (with respect to the cosmic time) as,
\begin{eqnarray}
 H(t) = \frac{1}{t_0}\left[\frac{2a_0n(t/t_0)}{\left(1 + a_0(t/t_0)^2\right)} + f_0\left(\frac{t-t_s}{t_0}\right)^{\alpha}\right]~~,
 \label{Hubble parameter-g}
\end{eqnarray}
and
\begin{eqnarray}
 \frac{dH}{dt} = \frac{1}{t_0^2}\left[\frac{2a_0n\left(1 - a_0(t/t_0)^2\right)}{\left(1 + a_0(t/t_0)^2\right)^2} 
 + \alpha f_0\left(\frac{t-t_s}{t_0}\right)^{\alpha-1}\right]~~,
 \label{der-H-g}
\end{eqnarray}
respectively. The above expression of $H(t)$ refers to different types of finite time singularity depending on the values of $\alpha$. In particular,
\begin{itemize}
 \item For $\alpha < -1$, a Type-I singularity appears at $t = t_s$, i.e the scale factor, the Hubble parameter and the derivative(s) of the 
 Hubble parameter simultaneously diverge at $t = t_s$. The divergence of the Hubble parameter and its first derivative indicates the divergence of 
 the effective energy density and the effective pressure respectively.
 
 \item The range $-1 < \alpha < 0$ leads to a Type-III singularity at $t = t_s$, i.e the scale factor tends to a finite value, while the Hubble parameter and 
 its derivative(s) diverge at $t = t_s$.
 
 \item With $0 < \alpha < 1$, a Type-II singularity appears at $t = t_s$. In this case, the scale factor and $H(t)$ tend to a finite value, 
 while $\dot{H}(t)$ (and also the higher derivatives) diverges at $t = t_s$.
 
 \item With $\alpha > 1$ and non-integer, 
 a Type-IV singularity appears at $t = t_s$, for which, the scale factor, $H(t)$ and $\dot{H}(t)$ 
 tend to a finite value at $t = t_s$, however the higher derivatives of the Hubble parameter diverge at the singularity point. Clearly in the case of 
 Type-IV singularity, the effective energy density and the effective pressure are finite at the singularity time. 
\end{itemize}

Therefore among these finite time singularities, the Type-IV singularity is the most mild singularity, 
and in the present context, we want to investigate the possible effects 
of such a mild singularity in an otherwise non-singular bounce cosmology. For this purpose, we take $\alpha > 1$. 
Here it deserves mentioning that the parameter $\alpha$ satisfies the following conditions -- (a) $\alpha$ should be greater than unity in order to have a Type-IV singularity 
at a finite time, (b) $\alpha$ should be of the form like $\alpha = \frac{2p+1}{2q+1}$ where $p$ and $q$ are positive integers, and (c) we 
consider the real root of $(-1)^{\alpha}$ (note that $(-1)^{\alpha}$ has two complex branches and one real negative), 
in particular, $(-1)^{\alpha} = -1$. The last two conditions ensure the requirement that 
both the scale factor and the Hubble parameter (see Eq.(\ref{scale factor-g}) and Eq.(\ref{Hubble parameter-g})) 
are real valued functions during the entire cosmic range in the present context. In particular, owing to such form of 
$\alpha$, the term $(t-t_s)^{\alpha+1}$ present in the scale factor becomes real-positive during $t>t_s$ and is real-negative during 
$t<t_s$. Therefore complex values of the scale factor are avoided with the aforementioned choice of $\alpha$. 
In this regard, we would like to mention that similar kind of scale factor has been considered in earlier literatures, see 
\cite{Odintsov:2015ynk,Odintsov:2021yva,Nojiri:2022xdo}.\\ 

Here we would like to mention that the cosmological evolution predicted from the scale factor of Eq.(\ref{scale factor-g}) can indeed 
be realized in the present 
context of Gauss-Bonnet (GB) theory of gravity with suitable forms of the scalar field potential and the GB coupling function. Integrating 
both sides of Eq.(\ref{constraint on coupling}), we get
\begin{eqnarray}
 \dot{h} = h_0a(t)~~,
 \label{new1}
\end{eqnarray}
where $h_0$ is a constant having mass dimension $=[-1]$. By using the above expression of $\dot{h}$, 
we evaluate the scalar field potential ($\tilde(V)(\chi)$) from Eq.(\ref{final form1}) and the Lagrange multiplier ($\lambda(t)$) 
from Eq.(\ref{final form2}):
\begin{eqnarray}
 \tilde{V}(\chi)&=&\left(2\dot{H} + 3H^2\right)\left(\frac{1}{\kappa^2} - 8h_0a(t)H(t)\right)\bigg|_{t-t_b = \chi/\mu^2}~~,\nonumber\\
\lambda(t)&=&\frac{2\dot{H}}{\mu^4}\left(\frac{1}{\kappa^2} + 8h_0a(t)H(t)\right)~~,
 \label{new2}
\end{eqnarray}
where $a(t)$, $H(t)$ and $\dot{H}(t)$ are given above in Eq.(\ref{scale factor-g}), Eq.(\ref{Hubble parameter-g}) and Eq.(\ref{der-H-g}) respectively. 
Since we are dealing with $\alpha > 1$ in order to have a Type-IV singularity at a finite time, the Hubble parameter and its first derivative 
are regular even at the time of singularity, which in turn ensures the regular behaviour of 
both the scalar field potential and the Lagrange multiplier during the entire cosmic evolution of the universe. 
It may be noted that both the scalar field potential and the Lagrange multiplier contain $\dot{H}$, and thus their derivatives diverge 
at the Type-IV singularity at $t=t_s$.\\

In a bouncing universe, the universe initially contracts where the Hubble parameter is negative, and after it bounces off, the universe 
enters to an expanding phase when the Hubble parameter becomes positive. Therefore at the bouncing point, the Hubble parameter satisfies 
the conditions: $H = 0$ and $\dot{H}>0$ respectively. Following, we will examine whether the 
scale factor of Eq.~(\ref{scale factor-g}) leads to a bouncing universe, and for this purpose, we separately consider $t_s < 0$, $t_s > 0$ and 
$t_s = 0$ respectively. 

\begin{enumerate}
 \item \textbf{\underline{For $t_s < 0$}}: Here we take $t_s = -|t_s|$. 
 Therefore in the cosmic regime $-\infty < t < -|t_s|$, both the terms of $H(t)$ in Eq.~(\ref{Hubble parameter-g}) are negative and thus there 
 is no possibility to have $H(t) = 0$ (or equivalently, a bounce) in this regime. However during $-|t_s| < t < 0$, the first term of $H(t)$ is negative, 
 while the second term containing $f_0$ becomes positive. Thus there is a possibility for $H(t) = 0$ during 
 $-|t_s| < t < 0$, which may lead to a bouncing universe. In particular, during the negative cosmic time, i.e for $-\infty < t < 0$, the 
 Hubble parameter from Eq.~(\ref{Hubble parameter-g}) can be written as,
\begin{eqnarray}
 H(t) = \frac{1}{t_0}\left[-\frac{2a_0n(|t|/t_0)}{\left(1 + a_0(|t|/t_0)^2\right)} + f_0\left(\frac{-|t|+|t_s|}{t_0}\right)^{\alpha}\right] 
 = -H_1(t) + H_2(t)~~(\mathrm{say})~~,
 \label{Hubble parameter-g-item1a}
 \end{eqnarray}
 where $H_1(t)$ and $H_2(t)$ are the first and the second terms respectively, sitting in the right hand side of the above expression. The 
 evolutions of $H_1(t)$ and $H_2(t)$, and their comparisons are shown in the following Table-[\ref{Table-com}].
 
  \begin{table}[t]
  \centering
%  \resizebox{\columnwidth}{1.0 cm}{%
  \begin{tabular}{|c|c|}
   \hline 
  $Evolution~of~H_1(t)$  & $Evolution~of~H_2(t)$\\
  \hline
   \hline\\
   $H_1(t) = 0$ at $t\rightarrow -\infty$ & $H_2(t)$ diverges to $-\infty$ at $t\rightarrow -\infty$ \\
   \hline\\
   $H_1(t)$ has a maximum within $-\infty < t < 0$ & $H_2(t)$ is an increasing function during $-\infty < t < 0$. \\
   \hline\\
   $H_1(t) = 0$ at $t = 0$ & $H_2(t) = 0$ at $t = -|t_s|$ and, $H_2(t) = f_0|t_s|^{\alpha}$ at $t = 0$. \\
   \hline
   \hline
  \end{tabular}%
%  }
  \caption{Comparison between the evolutions of $H_1(t)$ and $H_2(t)$ during $t \leq 0$.}
  \label{Table-com}
 \end{table}
 
 Therefore $H_1(t)$ starts from zero at the distant past and having a maximum within $-\infty < t < 0$, it again reaches to zero at $t = 0$. 
 On the other hand, $H_2(t)$ seems to be an increasing function during $-\infty < t < 0$, in particular, $H_2(t) \rightarrow -\infty$ at the distant past 
 and experiences a zero crossing at $t = -|t_s|$. These clearly argue that there exists a time (say, $t_b$) within 
 $-|t_s| < t < 0$ when $H(t_b) = 0$, i.e from Eq.~(\ref{Hubble parameter-g-item1a}),
 \begin{eqnarray}
  \frac{2a_0n(|t_b|/t_0)}{\left(1 + a_0(|t_b|/t_0)^2\right)} = f_0\left(\frac{-|t_b|+|t_s|}{t_0}\right)^{\alpha}~~.\label{bounce-global-item1}
 \end{eqnarray}
Thus as a whole,
\begin{itemize}
 \item $H_1(t) > H_2(t)$, or equivalently $H(t) < 0$, during $t < t_b$.
 
 \item $H_1(t) = H_2(t)$, or equivalently $H(t) = 0$, during $t = t_b$.
 
 \item $H_1(t) < H_2(t)$, or equivalently $H(t) > 0$, during $t > t_b$.
\end{itemize}
This implies that $t = t_b$ ($= - |t_b|$) is the bouncing time which can be determined from Eq.~(\ref{bounce-global-item1}). A closed solution 
of $t_b$ may not be possible, however, we numerically obtain the solution of $t_b$ from Eq.~(\ref{bounce-global-item1}) for a suitable set 
of parameter values. In particular, for $f_0 = 1$, $n = 0.3$, $a_0 = 4$, $\alpha = \frac{5}{3}$, $t_0 = 1\mathrm{By}$ and $t_s = -1\mathrm{By}$ 
(where 'By' stands for Billion year), we obtain $t_b = -0.31\mathrm{By}$. 
In general, Eq.~(\ref{bounce-global-item1}) clearly indicates that for $t_s < 0$, the Type-IV singularity occurs before the instant of the bounce, i.e 
$t_s < t_b$ or $|t_s| > |t_b|$. 

\item \textbf{\underline{For $t_s > 0$}}: Performing the same procedure as we have done in the previous case, we argue that for $t_s > 0$, the 
scale factor of Eq.~(\ref{scale factor-g}) leads to a bouncing universe and the bounce happens within $0 < t < t_s$. If the instant of bounce 
is denoted by $t_b$, then from Eq.~(\ref{Hubble parameter-g}), we obtain,
\begin{eqnarray}
 \frac{2a_0n(t_b/t_0)}{\left(1 + a_0(t_b/t_0)^2\right)} = f_0\left(\frac{t_s - t_b}{t_0}\right)^{\alpha}~~.\label{bounce-global-item2}~~.
\end{eqnarray}
Once again, one may numerically solve $t_b$ from the above algebraic equation with a suitable set of parameter values. In particular, for 
$f_0 = 1$, $n = 0.3$, $a_0 = 4$, $\alpha = \frac{5}{3}$, $t_0 = 1\mathrm{By}$ and $t_s = 1\mathrm{By}$, the bounce time comes as 
$t_b = 0.31\mathrm{By}$. From Eq.~(\ref{bounce-global-item2}), it is clear that for $t_s > 0$, the Type-IV singularity occurs after the bounce, 
i.e $t_s > t_b$. 

\item \textbf{\underline{For $t_s = 0$}}: The expression of $H(t)$ of Eq.~(\ref{Hubble parameter-g}) reveals that for 
$t_s = 0$, the universe experiences a bounce at $t = 0$. Thus the condition $t_s = 0$ leads to the fact that the Type-IV singularity occurs 
at the instant of bounce.\\
 
 \end{enumerate}

 Therefore in all the three cases, the scale factor we consider in Eq.~(\ref{scale factor-g}) leads to a bouncing universe along with a Type-IV 
 singularity for $\alpha > 1$. Here it deserves mentioning that the presence of the Type-IV singularity affects the time of bounce. In particular --
 (1) if the singularity occurs at $t_s < 0$, then the bounce shows at some negative instant of time, and moreover, it tuns out that 
 the singularity occurs before the bounce happens. (2) The case $t_s > 0$ leads to the bounce at some positive time and consequently, the singularity 
 occurs after the bounce, and finally, (3) for $t_s = 0$, the bounce and the Type-IV singularity appears at the same instant of time $t = 0$.\\
 
\begin{figure}[!h]
\begin{center}
\centering
\includegraphics[width=3.2in,height=2.5in]{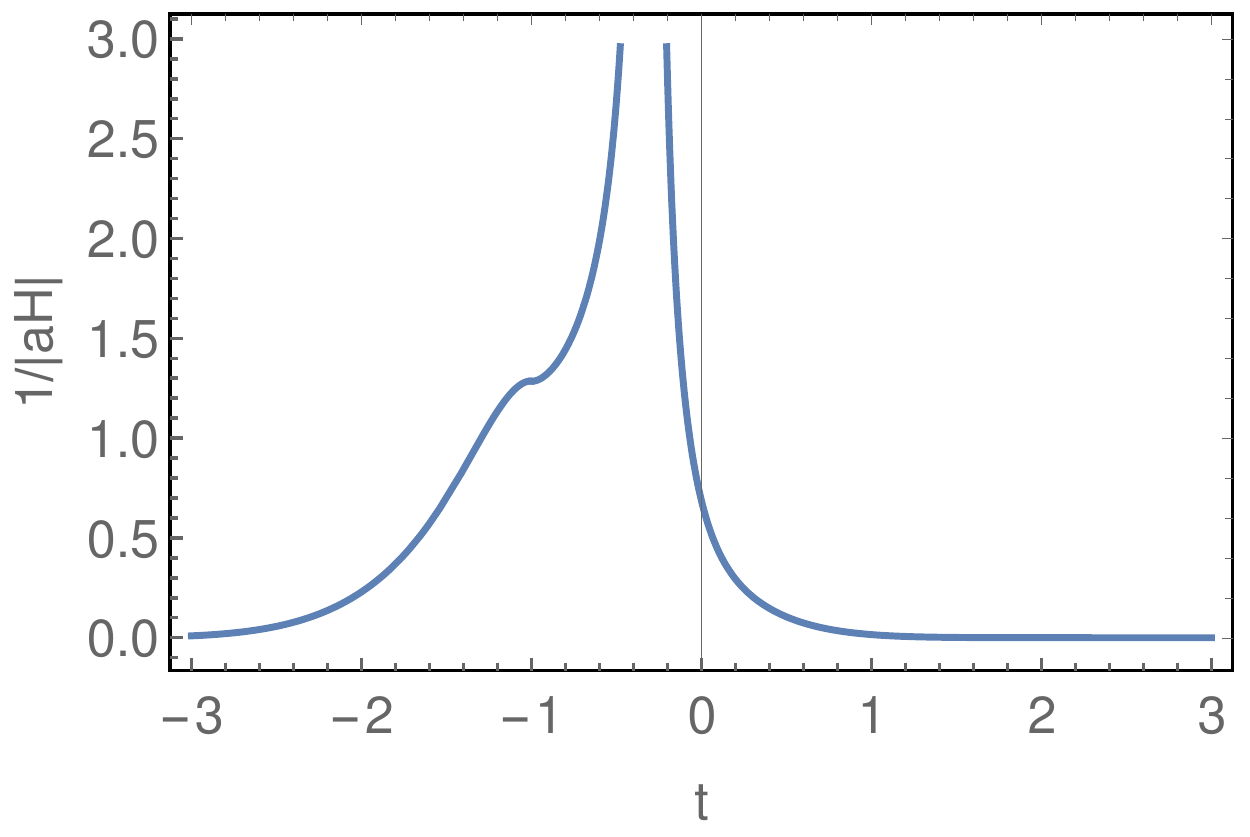}
\includegraphics[width=3.2in,height=2.5in]{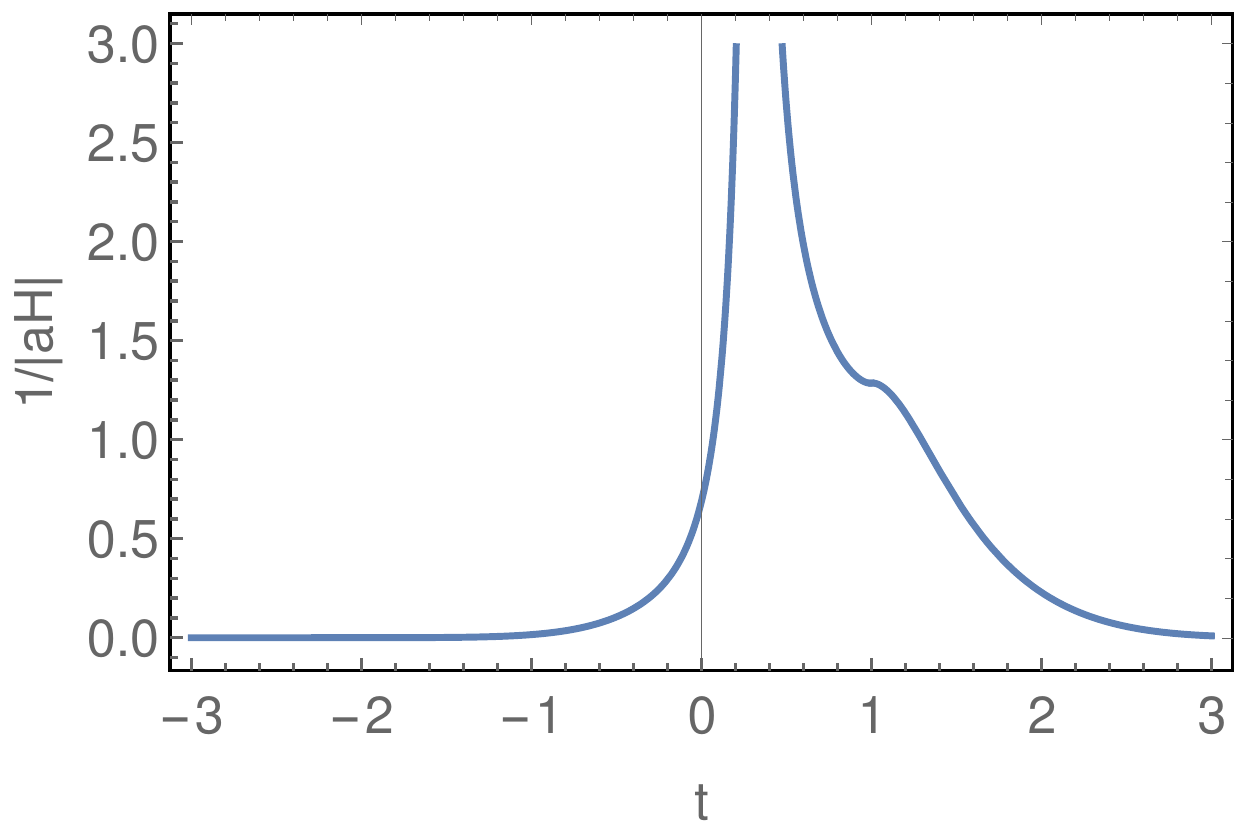}
\caption{$r_h(t)$ vs. $t$. Here we take $n = 0.3$, $a_0 = 4$, $\alpha = 5/3$ and $f_0 = 1$. 
We will show that the observable quantities, in the case when the Type-IV singularity globally affects the spacetime, 
are not compatible with the Planck data. Hence in order to present the plots, we consider a set of values of the parameters by keeping the folowing points 
in mind -- (a) the value of $\alpha$ should satisfy the form of $\alpha = \frac{2p+1}{2q+1}$ as we have mentioned after eq.(\ref{scale factor-g}), and 
(b) the parameter $n$ should be less than $\frac{1}{2}$, so that the comoving Hubble radius predicetd from $a_1(t)$ 
diverges to infinity at the distant past and realizes the effects of the Type-IV singularity on the bouncing dynaimcs. Thus for example, we consider 
$n = 0.3$, $a_0 = 4$, $\alpha = 5/3$ and $f_0 = 1$ -- which, in fact, leads to the viability of the bounce model 
when the Type-IV singularity locally affects 
the spacetime, see Sec.[\ref{sec-local}]. 
The left and right plots correspond to $t_s = -1\mathrm{By}$ and $t_s  = 1\mathrm{By}$ 
respectively. $t_0$ is taken to be $1\mathrm{By}$ to make all the time coordinates in the unit 
of Billion year (By). Therefore in the left plot $t_s < t_b = -0.31\mathrm{By}$, and for the right plot $t_s > t_b = 0.31\mathrm{By}$.}
\label{plot-rh1}
\end{center}
\end{figure}

\begin{figure}[!h]
\begin{center}
\centering
\includegraphics[width=3.5in,height=2.5in]{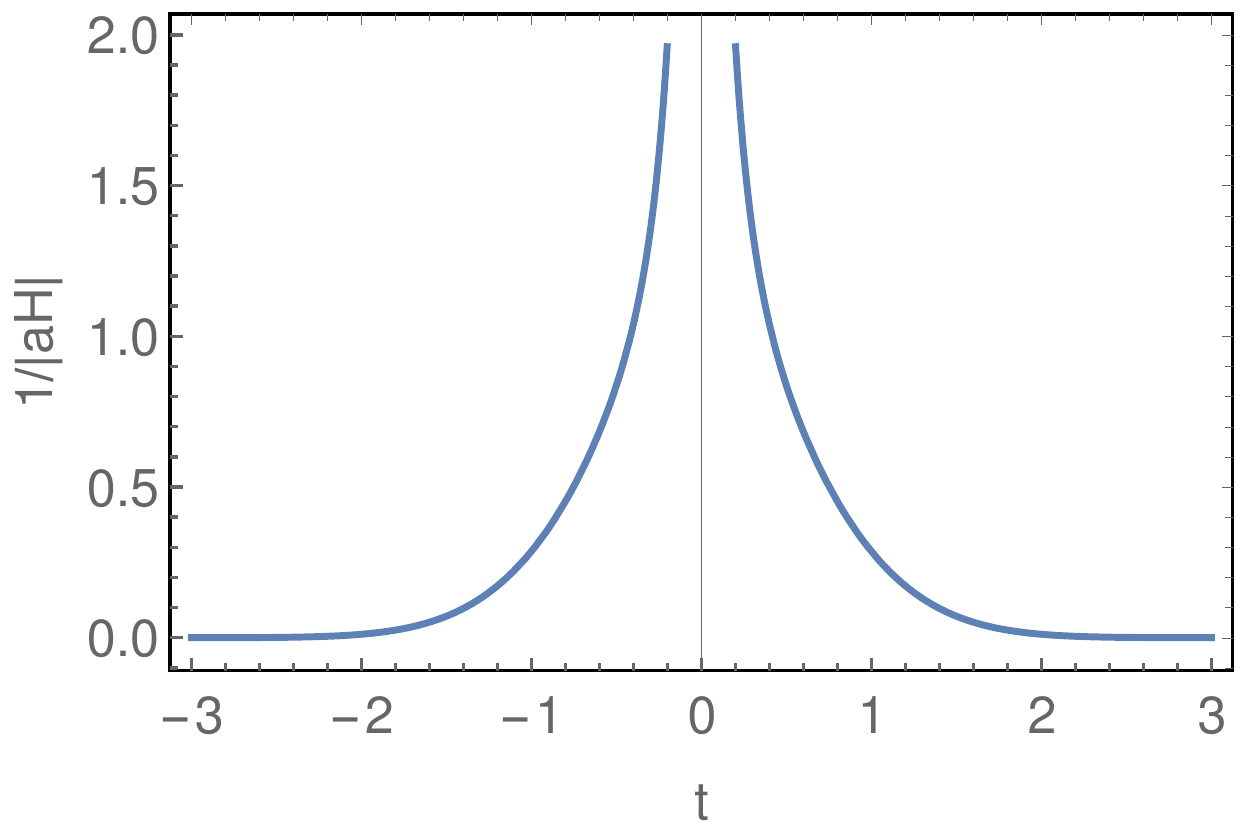}
\caption{$r_h(t)$ vs. $t$ for $t_s = 0$. The parameter values are considered to be same as in the earlier figure. In this case $t_s = t_b$.}
\label{plot-rh2}
\end{center}
\end{figure}

 The comoving Hubble radius is defined by $r_h = 1/|aH|$ which, by definition, diverges at bounce. By using 
 Eq.~(\ref{scale factor-g}) and Eq.~(\ref{Hubble parameter-g}), we give the plots for $r_h$ vs. $t$ for the above three cases, 
 see Fig.[\ref{plot-rh1}] and Fig.[\ref{plot-rh2}].
 
 The left plot of the Fig.[\ref{plot-rh1}] corresponds to the case $t_s < 0$, which clearly depicts that $r_h$ diverges 
 or equivalently the bounce happens at a $t < 0$, and moreover, the Type-IV singularity is found to occur before the bounce 
 (see the discussion in the caption of the figure). The other plots of $r_h(t)$ are also in accordance with the arguments mentioned above. 
 One important point may be noticed that irrespective of whether $t_s < 0$ or $t_s > 0$ or $t_s = 0$, the comoving Hubble radius asymptotically goes to zero 
 at both sides of the bounce. Actually $r_h$ at the distant past or at the distant future comes with the following form,
 \begin{eqnarray}
  \lim_{t\rightarrow \pm\infty}~r_h = \left|\frac{1}{|t|^{2n+\alpha}}\exp{\left[-\frac{f_0}{\alpha+1}|t|^{\alpha+1}\right]}\right|~~,\label{hubble radius-g}
 \end{eqnarray}
which is independent of $t_s$, and clearly demonstrates that $r_h$ asymptotically vanishes at $t\rightarrow \pm\infty$. Moreover, this argument 
holds for any value of the parameter $n$. In a bouncing universe, the primordial 
perturbation modes generate either near the bounce or far before the bounce depending on the asymptotic evolution of comoving Hubble radius. In the 
present context, we find that $r_h\rightarrow 0$ at $t\rightarrow -\infty$, which suggests that the perturbation modes generate near the bounce when all 
the modes lie within the sub-Hubble regime. Here it may be mentioned that for $f_0 = 0$ (i.e in absence of the Type-IV singularity), the scale factor 
is described by $a(t) = \left(1 + a_0t^2\right)^n$ and consequently $r_h$ asymptotically diverges to infinity (for $n < \frac{1}{2}$). 
Therefore it turns out that for $f_0 = 0$, the primordial perturbation modes generate far away from the bounce in the deep contracting phase, unlike to 
the case where $f_0 \neq 0$. Actually for $f_0 \neq 0$, the presence of the Type-IV singularity globally affects the evolution of the Hubble parameter 
(compared to the case where the singularity is absent), which in turn leads to the comoving Hubble radius 
tending to zero at $t \rightarrow \pm\infty$ and makes the generation era of the perturbation modes near the bounce.

\subsection*{Cosmological perturbation}

Since the perturbation modes generate and cross the horizon near the bounce, we are interested to examine the evolution of scalar and tensor perturbations 
near the bouncing phase. For this purpose, the useful quantities are the scale factor and the Hubble parameter near $t = t_b$ (recall, $t_b$ is 
the time of the bounce), and they are given by:
\begin{eqnarray}
 a(t)&=&a_\mathrm{b}\left[1 + \dot{H}_\mathrm{b}\frac{\left(t - t_b\right)^2}{2}\right]~~,\nonumber\\
 H(t)&=&\dot{H}_\mathrm{b}\left(t - t_b\right)~~,\label{Hubble nb}
\end{eqnarray}
where we use the Taylor series expansion around $t = t_b$. 
Thereby using Eq.~(\ref{scale factor-g}) and Eq.~(\ref{Hubble parameter-g}), we get,
\begin{eqnarray}
 a_\mathrm{b}&=&\left(1+a_0\left(\frac{t_b}{t_0}\right)^2\right)^{n}\times 
 \exp{\left[\frac{f_0}{(\alpha+1)}\left(\frac{t_b-t_s}{t_0}\right)^{\alpha+1}\right]}~~,\nonumber\\
 \dot{H}_\mathrm{b}&=&\frac{1}{t_0^2}\left\{\frac{2a_0n\left[1 - a_0\left(t_b/t_0\right)^2\right]}
 {\left[1 + a_0\left(t_b/t_0\right)^2\right]^2} + \alpha f_0\left(\frac{t_b - t_s}{t_0}\right)^{\alpha -1}\right\}~~.\label{at bounce expression}
\end{eqnarray}
The Gauss-Bonnet coupling function in the present context obeys $\ddot{h} = \dot{h}H$ from the requirement 
of compatibility with the GW170817 event. Integrating $\ddot{h} = \dot{h}H$ with respect to the cosmic time, one gets,
\begin{eqnarray}
 \dot{h} = \frac{1}{\kappa}a(t) = \frac{a_\mathrm{b}}{\kappa}\left[1 + \dot{H}_\mathrm{b}\frac{\left(t - t_b\right)^2}{2}\right]~~,\label{GB nb}
\end{eqnarray}
where the integration constant is taken as $1/\kappa$ from the dimensional analysis. The above forms of $a(t)$ and $H(t)$ along with the 
solution of $\chi(t) = \mu^2(t-t_b)$ lead to the scalar field potential and the Lagrange multiplier from Eq.~(\ref{final form2}) as,
\begin{eqnarray}
 \tilde{V}(\chi)&=&\frac{2}{\kappa^2}\dot{H}_\mathrm{b} - \frac{16a_\mathrm{b}(\dot{H}_\mathrm{b})^2}{\kappa}\left(\frac{\chi}{\mu^2}\right) 
 + \frac{3(\dot{H}_\mathrm{b})^2}{\kappa^2}\left(\frac{\chi}{\mu^2}\right)^2 
 - \frac{32a_\mathrm{b}(\dot{H}_\mathrm{b})^3}{\kappa}\left(\frac{\chi}{\mu^2}\right)^3 
 - \frac{12a_\mathrm{b}(\dot{H}_\mathrm{b})^4}{\kappa}\left(\frac{\chi}{\mu^2}\right)^5~~,\label{potential nb}\\
 \mu^4\lambda(t)&=&\frac{2}{\kappa^2}\dot{H}_\mathrm{b} + \frac{16a_\mathrm{b}(\dot{H}_\mathrm{b})^2}{\kappa}\left(t - t_b\right) 
 + \frac{8a_\mathrm{b}(\dot{H}_\mathrm{b})^3}{\kappa}\left(t-t_b\right)^3~~,\label{LM nb}
\end{eqnarray}
respectively. In the context of Gauss-Bonnet theory of gravity, we may introduce an effective potential for the scalar field as,
\begin{eqnarray}
 V_\mathrm{eff}(\chi) = \tilde{V}(\chi) + 24h(\chi)\left[H^4 + H^2\dot{H}\right]\nonumber
\end{eqnarray}
which, due to Eq.~(\ref{Hubble nb}) along with $\chi(t) = \mu^2(t-t_b)$, can be expressed around $t = t_b$ as,
\begin{eqnarray}
 V_\mathrm{eff}(\chi) = \tilde{V}(\chi) + 24h(\chi)\left[\dot{H}_\mathrm{b}^4\left(\frac{\chi}{\mu^2}\right)^4 
 + \dot{H}_\mathrm{b}^3\left(\frac{\chi}{\mu^2}\right)^2\right]~~.\label{effective potential nb}
\end{eqnarray}
We will use the above expressions to examine the evolution of scalar and tensor perturbations, and consequently in determination of various observable 
quantities like the scalar spectral index and tensor-to-scalar ratio.

\subsubsection*{Scalar perturbation}
The scalar perturbation over the FRW metric in the longitudinal gauge is,
\begin{eqnarray}
 ds^2 = a^2(\eta)\bigg[\big(1 + 2\Psi\big)d\eta^2 - \big(1 - 2\Psi\big)\delta_{ij}dx^idx^j\bigg]
 \label{sp1}
\end{eqnarray}
where $\Psi(\eta,\vec{x})$ is the scalar perturbation variable and $\eta$ is the conformal time coordinate. Here it may be mentioned that 
the background evolution has no anisotropic stress, so we work with one scalar perturbation variable, namely $\Psi(\eta,\vec{x})$.

The scalar field is perturbed as,
\begin{eqnarray}
 \chi(\eta,\vec{x}) = \chi_0(\eta) + \delta\chi(\eta,\vec{x})
 \label{sp2-g}
\end{eqnarray}
where $\chi_0$ is the background evolution of the scalar field, and given by $\chi_0(t) = \mu^2(t-t_b)$ in terms of cosmic time with 
$t_b$ being the instant of bounce. The scalar perturbation equations up-to the leading order in the longitudinal gauge are given by 
\cite{Brandenberger:2003vk},
\begin{eqnarray}
\nabla^2\Psi - 3\mathcal{H}\Psi' - 3\mathcal{H}\Psi&=&\frac{\kappa^2}{2}a^2\delta T^0_0\nonumber\\
\big(\Psi' + \mathcal{H}\Psi\big)_{,i}&=&\frac{\kappa^2}{2}a^2\delta T^0_i\nonumber\\
\bigg[\Psi'' + 3\mathcal{H}\Psi' + \big(2\mathcal{H}' + \mathcal{H}^2\big)\Psi\bigg]\delta^i_j&=&-\frac{\kappa^2}{2}a^2\delta T^i_j
\label{sp3-g}
\end{eqnarray}
where $\mathcal{H} = \frac{a'}{a}$ is the Hubble parameter in the $\eta$ coordinate and $\delta T_{\mu\nu}$ denotes 
the variation of energy-momentum tensor due to the perturbation of the spacetime metric and the scalar field, mentioned above. The variation of 
matter energy-momentum tensor in the present context comes with the following expressions,
\begin{eqnarray}
 \delta T^0_0&=&\frac{1}{a^2}\bigg[\lambda(t)\chi_0'\delta\chi' + a^2V_\mathrm{eff}'(\chi_0)\delta\chi\bigg]\nonumber\\
 \delta T^0_i&=&\frac{1}{a^2}\partial_{i}\bigg[\lambda(t)\chi_0'\delta\chi\bigg]\nonumber\\
 \delta T^{i}_{j}&=&-\frac{1}{a^2}\delta^i_j
 \bigg[\lambda(\Phi_0)\chi_0'\delta\chi' - a^2V_\mathrm{eff}'(\chi_0)\delta\chi\bigg]~~,
 \label{sp4-g}
\end{eqnarray}
where we use Eq.~(\ref{sp2-g}), and $V_\mathrm{eff}$ is obtained earlier in Eq.~(\ref{effective potential nb}). 
Plugging back the above expressions of $\delta T_{\mu\nu}$ into Eq.~(\ref{sp3-g}) yields the following set of equations:
\begin{eqnarray}
 \nabla^2\Psi - 3\mathcal{H}\Psi' - 3\mathcal{H}^2\Psi&=&\frac{\kappa^2}{2}
 \bigg[\lambda(\Phi_0)\chi_0'\delta\chi' + a^2V_\mathrm{eff}'(\chi_0)\delta\chi\bigg]\nonumber\\
 \Psi' + \mathcal{H}\Psi&=&\frac{\kappa^2}{2}\lambda(t)\chi_0'\delta\chi\nonumber\\
 \Psi'' + 3\mathcal{H}\Psi' + \big(2\mathcal{H}' + \mathcal{H}^2\big)\Psi&=&\frac{\kappa^2}{2}
 \bigg[\lambda(\Phi_0)\chi_0'\delta\chi' - a^2V_\mathrm{eff}'(\chi_0)\delta\chi\bigg]
 \label{sp5-g}
\end{eqnarray}
respectively. The second equality of Eq.~(\ref{sp5-g}) helps to extract $\delta \chi$ in terms of $\Psi$ and $\Psi'$, using which with the other two 
equalities, one gets the following equation for $\Psi(\eta,\vec{x})$,  
\begin{eqnarray}
 \Psi'' - \nabla^2\Psi + 6\mathcal{H}\Psi' + \big(2\mathcal{H}' + 4\mathcal{H}^2\big)\Psi = 
 -2a^2V_\mathrm{eff}'(\chi_0)\left(\frac{\Psi' + \mathcal{H}\Psi}{\lambda(t)\chi_0'}\right)
 \label{sp6-g}
\end{eqnarray}
Clearly, $\Psi$ depends on the background evolution through $\mathcal{H}$ and $\chi_0$ which have been determined in the previous section in terms of the 
cosmic time (t). So it will be more useful if we transform the above equation into cosmic time, for which, we need the following relations: 
\begin{eqnarray}
 \Psi' = a\dot{\Psi}~~~~~~~~~~~~~~~~~~~~~\rm{and}~~~~~~~~~~~~~~~~~~~~\Psi'' = a^2\ddot{\Psi} + a^2H\dot{\Psi}~~.
 \nonumber
\end{eqnarray}
Accordingly Eq.~(\ref{sp6-g}) is given by,
\begin{eqnarray}
 \ddot{\Psi} - \frac{1}{a^2}\nabla^2\Psi + \bigg[7H + \frac{2V_\mathrm{eff}'(\chi_0)}{\mu^2\lambda}\bigg]\dot{\Psi} 
 + \bigg[2\dot{H} + 6H^2 + 2H\left(\frac{2V_\mathrm{eff}'(\chi_0)}{\mu^2\lambda}\right)\bigg]\Psi = 0
 \label{sp8-g}
\end{eqnarray}
where $H = \frac{\dot{a}}{a}$ is the Hubble parameter in cosmic time. As we have mentioned earlier, owing to the presence of the Type-IV 
singularity, the comoving Hubble radius asymptotically goes to zero at both sides of the bounce and thus the perturbation modes generate and cross the 
horizon near the bounce when all the modes lie within the Hubble radius. Therefore we intend to solve Eq.~(\ref{sp8-g}) near the bounce, and thus 
we use the near-bounce expressions of $H(t)$ and $V_\mathrm{eff}(\chi_0)$ from Eq.~(\ref{Hubble nb}) and Eq.~(\ref{effective potential nb}) 
respectively. Consequently we get,
\begin{eqnarray}
 \frac{V_\mathrm{eff}'(\chi_0)}{\mu^2\lambda} = 
-8\kappa a_\mathrm{b}\dot{H}_\mathrm{b} + \left\{3\dot{H}_\mathrm{b} + \left(8\kappa a_\mathrm{b}\dot{H}_\mathrm{b}\right)^2\right\}(t-t_b) 
 \label{sp9-g}
\end{eqnarray}
where we retain up-to the leading order in $t-t_b$. With the above expression, Eq.~(\ref{sp8-g}) can be written as, 
\begin{eqnarray}
 \ddot{\Psi} - \nabla^2\Psi + \left[-2p + q(t-t_b)\right]\dot{\Psi} 
 + \left[2\dot{H}_\mathrm{b} - 2p\dot{H}_\mathrm{b}(t-t_b)\right]\Psi(\vec{x},t) = 0~~,
 \label{sp10-g}
\end{eqnarray}
with $p$ and $q$ are given by,
\begin{eqnarray}
 p = 8\kappa a_\mathrm{b}\dot{H}_\mathrm{b}~~~~~\mathrm{and}~~~~~
 q = 2p^2 + 13\dot{H}_b
 \label{p-q}
\end{eqnarray}
respectively. The Fourier transformation of Eq.~(\ref{sp10-g}) is,
\begin{eqnarray}
 \ddot{\Psi}_k + \left[-2p + q(t-t_b)\right]\dot{\Psi}_k 
 + \left[\left(k^2 + 2\dot{H}_\mathrm{b}\right) - 2p\dot{H}_\mathrm{b}(t-t_b)\right]\Psi_k = 0~~,
 \label{sp11-g}
\end{eqnarray}
where $\Psi_k(t)$ is the Fourier transformed variable of $\Psi(t,\vec{x})$. Eq.~(\ref{sp11-g}) has the following solution 
for $\Psi_k(t)$:
\begin{eqnarray}
 \Psi_k(t) = C(k)~\exp{\left[2p\left(1-\frac{\dot{H}_\mathrm{b}}{q}\right)(t-t_b)\right]}\times
 H_\mathrm{s}\left[\sqrt{2}\left(\frac{p}{q^{3/2}}\right)\left(2\dot{H}_\mathrm{b} - q\right) + \sqrt{\frac{q}{2}}(t-t_b)\right]~~,
 \label{sp12-g}
\end{eqnarray}
with $H_\mathrm{s}[x]$ is the Hermite polynomial having order $s$. The order of the Hermite polynomial in the above expression is given by,
\begin{eqnarray}
 s = -1 + \frac{k^2 + 2\dot{H}_\mathrm{b}}{q} + \frac{4p^2\dot{H}_\mathrm{b}\left(\dot{H}_\mathrm{b} - q\right)}{q^3}~~.\label{s}
\end{eqnarray}
Moreover $C(k)$ is the integration constant which can be determined from the Bunch-Davies condition given by 
$\lim_{\eta\rightarrow\eta_b}v_k(\eta) = \frac{1}{\sqrt{2k}}e^{-ik\eta}$, where $v_k(\eta)$ is the canonical scalar Mukhanov-Sasaki variable.
The Bunch-Davies condition is well justified from the fact that the perturbation modes near the bounce lie within the Hubble radius. 
The Bunch-Davies vacuum condition on the Mukhanov-Sasaki variable can be transformed into the corresponding condition on 
$\Psi_k(t)$ through the following relation \cite{Brandenberger:2003vk},
\begin{eqnarray}
 \lim_{t \rightarrow t_b}\Psi_k(t) = \frac{\kappa^2}{2k^2}~\lim_{t \rightarrow t_b}\left[\sqrt{\lambda(t)}~\dot{\chi}_0\right]v_k'(\eta) 
 = \frac{i\kappa^2}{2\sqrt{2}k^{3/2}}\left(\mu^2\sqrt{\lambda(t_b)}\right)~~.
 \label{sp13-g}
\end{eqnarray}
where we use $\dot{\chi} = \mu^2$. Due to $\lambda(t_b) = 2\dot{H}_\mathrm{b}/(\mu^4\kappa^2)$ from Eq.~(\ref{LM nb}), the above equation 
can be equivalently written as,
\begin{eqnarray}
 \lim_{t \rightarrow t_b}\Psi_k(t) = \frac{i\kappa\sqrt{\dot{H}_\mathrm{b}}}{2k^{3/2}}~~.
 \label{sp14-g}
\end{eqnarray}
Consequently the integration constant $C(k)$ gets the following form,
\begin{eqnarray}
 C(k) = \frac{i\kappa\sqrt{\dot{H}_\mathrm{b}}}{2k^{3/2}}~\bigg\{\frac{1}
 {H_\mathrm{s}\left[\sqrt{2}\left(\frac{p}{q^{3/2}}\right)\left(2\dot{H}_\mathrm{b} - q\right)\right]}\bigg\}~~.
 \nonumber
\end{eqnarray}
Accordingly the solution of the scalar perturbation from Eq.~(\ref{sp12-g}) turns out to be,
\begin{eqnarray}
 \Psi_k(t) = \frac{i\kappa\sqrt{\dot{H}_\mathrm{b}}}{2k^{3/2}}~\exp{\left[2p\left(1-\frac{\dot{H}_\mathrm{b}}{q}\right)(t-t_b)\right]}
 \left\{\frac{H_\mathrm{s}\left[\sqrt{2}\left(\frac{p}{q^{3/2}}\right)\left(2\dot{H}_\mathrm{b} - q\right) + \sqrt{\frac{q}{2}}(t-t_b)\right]}
 {H_\mathrm{s}\left[\sqrt{2}\left(\frac{p}{q^{3/2}}\right)\left(2\dot{H}_\mathrm{b} - q\right)\right]}\right\}
 \label{sp15-g}
\end{eqnarray}
with $p$ and $q$ being shown in Eq.~(\ref{p-q}). 
Consequently the scalar power spectrum for $k$-th mode is determined as,
\begin{eqnarray}
 P_{\Psi}(k,t)&=&\frac{k^3}{2\pi^2}\bigg|\Psi_k(t)\bigg|^2\nonumber\\
  &=&\frac{\kappa^2\dot{H}_\mathrm{b}}{8\pi^2}~\exp{\left[4p\left(1-\frac{\dot{H}_\mathrm{b}}{q}\right)(t-t_b)\right]}
 \left|\frac{H_\mathrm{s}\left[\sqrt{2}\left(\frac{p}{q^{3/2}}\right)\left(2\dot{H}_\mathrm{b} - q\right) + \sqrt{\frac{q}{2}}(t-t_b)\right]}
 {H_\mathrm{s}\left[\sqrt{2}\left(\frac{p}{q^{3/2}}\right)\left(2\dot{H}_\mathrm{b} - q\right)\right]}\right|^2~~.
 \label{sp16-g}
\end{eqnarray}
We are interested to determine the observable quantities like the scalar spectral index ($n_s$) and the tensor-to-scalar ratio ($r$), and 
will examine the possible effects of the Type-IV singularity on such observable indices. The horizon crossing condition for the 
$k$-th mode is $k = |aH|$, where $k \sim 0.05\mathrm{Mpc}^{-1}$, i.e we intend to calculate $n_s$ and $r$ over the large scale modes. 
The occurrence of the Type-IV singularity in the present context leads to 
the large scale modes crossing the horizon near the bounce, and thus, by using 
the near-bounce expression of $H(t)$ (see Eq.~(\ref{Hubble nb})) the horizon crossing condition can be written as,
\begin{eqnarray}
 t_h - t_b = \left(\frac{k}{a_\mathrm{b}\dot{H}_\mathrm{b}}\right)~~,\label{hc-global1}
\end{eqnarray}
where $t_h$ symbolizes the horizon crossing instant of the $k$-th mode. Therefore the scalar power spectrum at the horizon crossing 
is given by,
\begin{eqnarray}
 P_{\Psi}(k,t_h)&=&\frac{k^3}{2\pi^2}\bigg|\Psi_k(t)\bigg|^2\nonumber\\
  &=&\frac{\kappa^2\dot{H}_\mathrm{b}}{8\pi^2}~\exp{\left[4p\left(1-\frac{\dot{H}_\mathrm{b}}{q}\right)
  \left(\frac{k}{a_\mathrm{b}\dot{H}_\mathrm{b}}\right)\right]}
 \left|\frac{H_\mathrm{s}\left[\sqrt{2}\left(\frac{p}{q^{3/2}}\right)\left(2\dot{H}_\mathrm{b} - q\right) + \sqrt{\frac{q}{2}}
 \left(\frac{k}{a_\mathrm{b}\dot{H}_\mathrm{b}}\right)\right]}
 {H_\mathrm{s}\left[\sqrt{2}\left(\frac{p}{q^{3/2}}\right)\left(2\dot{H}_\mathrm{b} - q\right)\right]}\right|^2~~.
 \label{sp17-g}
\end{eqnarray}
which, clearly depends on $k$ through the term 
containing $k/(a_\mathrm{b}\dot{H}_\mathrm{b})$ as well as through the factor $s$ (the order of the Hermite polynomial, see Eq.~(\ref{s})). In particular, 
the spectral tilt of the scalar power spectrum is defined by,
\begin{eqnarray}
 n_s = 1 + \frac{\partial\ln{\left[P_{\Psi}(k,t_h)\right]}}{\partial\ln{k}}~~.
 \label{sp18-g}
\end{eqnarray}
However before estimating the $n_s$, let us perform the tensor perturbation which is useful for the observable quantity, namely the 
tensor-to-scalar ratio.

\subsubsection*{Tensor perturbation}
The tensor perturbation variable satisfies the following equation,
\begin{eqnarray}
  \frac{1}{a(t)z_T^2(t)}\frac{d}{dt}\bigg[a(t)z_T^2(t)\dot{h}_{ij}\bigg] - \frac{1}{a^2}\partial_{l}\partial^{l}h_{ij} = 0
  \label{ten per eom1-g}
 \end{eqnarray}
 where $h_{ij}(t,\vec{x})$ is the tensor perturbation variable and $z_T^2$, in the context of Lagrange multiplier Gauss-Bonnet gravity, 
 is given by \cite{Hwang:2005hb,Noh:2001ia,Hwang:2002fp},
 \begin{eqnarray}
  z_T^2 = \frac{a^2}{2\kappa^2}\left[1 - 16\kappa^2\dot{h}H\right]~~.\label{tp1-g}
 \end{eqnarray}
Using Eq.~(\ref{Hubble nb}) and Eq.~(\ref{GB nb}), we determine $z_T^2$ as,
\begin{eqnarray}
 a(t)z_T^2(t) = \frac{a_\mathrm{b}^3}{2\kappa^2}\left[1 - 16\kappa a_\mathrm{b}\dot{H}_\mathrm{b}(t-t_b) 
 + \frac{1}{2}\dot{H}_\mathrm{b}(t-t_b)^2\right]~~.\label{tp2-g}
\end{eqnarray}
The Fourier transformed tensor perturbation variable is defined as $h_{ij}(t,\vec{x}) = \int d\vec{k}~\sum_{\gamma}\epsilon_{ij}^{(\gamma)}~
h_{(\gamma)}(\vec{k},t) e^{i\vec{k}.\vec{x}}$, where $\gamma = '+'$ and $\gamma = '\times'$ represent two polarization modes. Therefore the above 
form of $z_T^2$ along with Eq.~(\ref{ten per eom1-g}) leads to the tensor perturbed equation in terms of the Fourier transformed variable as follows,
 \begin{eqnarray}
  \ddot{h}_k - 2p\left[1 + 2p(t-t_b)\right]\dot{h}_k + k^2h_k = 0~~,
  \label{ten per eom2-g}
 \end{eqnarray}
 where we retain the terms up-to the leading order in $\mathcal{O}(t-t_b)$, and recall that $p = 8\kappa a_\mathrm{b}\dot{H}_\mathrm{b}$ (see 
 Eq.~(\ref{p-q})). Here it may be mentioned that both the 
tensor polarizations ($\times$ and $+$ modes) in the present context obey the same differential Eq.~(\ref{ten per eom2-g}), due to which, we do not put 
any polarization index in the tensor perturbation variable. However we will multiply by a factor of '2' in the final expression of the 
tensor power spectrum due to their equal contribution to the spectrum. Solving Eq.~(\ref{ten per eom2-g}), we get,
\begin{eqnarray}
 h_k(t) = D(k)\times H_\mathrm{\omega}\left[\frac{1}{\sqrt{2}} + \sqrt{2}p(t-t_b)\right]~~,
 \label{ten per sol1-g}
\end{eqnarray}
where $\omega$ is the order of the Hermite polynomial and given by,
\begin{eqnarray}
 \omega = \frac{k^2}{\left(16\kappa a_\mathrm{b}\dot{H}_\mathrm{b}\right)^2}~~.\label{omega}
\end{eqnarray}
Moreover the integration constant $D(k)$ can be determined from the Bunch-Davies vacuum state near the bounce when the relevant 
perturbation modes lie within the sub-Hubble regime. In particular, the Bunch-Davies vacuum state is defined by 
$\lim_{t\rightarrow t_b}\big[z_T(t)h_k(t)\big] = \frac{1}{\sqrt{2k}}$. Due to $z_T(t_b) = a_\mathrm{b}/(\sqrt{2}\kappa)$ from 
Eq.~(\ref{tp2-g}), the Bunch-Davies condition results to,
\begin{eqnarray}
 D(k) = \frac{\kappa}{a_\mathrm{b}\sqrt{k}}\left[\frac{1}{H_\mathrm{\omega}\left[1/\sqrt{2}\right]}\right]~~~.
 \label{ten per bc-g}
\end{eqnarray}
Accordingly the final solution of $h_k(t)$ turns out to be,
\begin{eqnarray}
 h_k(t) = \frac{\kappa}{a_\mathrm{b}\sqrt{k}}\times\left\{\frac{H_\mathrm{\omega}\left[\frac{1}{\sqrt{2}} + \sqrt{2}p(t-t_b)\right]}
 {H_\mathrm{\omega}\left[1/\sqrt{2}\right]}\right\}~~.
 \label{ten per sol2-g}
\end{eqnarray}
Consequently the tensor power spectrum for the $k$-th mode is given by,
\begin{eqnarray}
 P_{h}(k,t)&=&\frac{k^3}{2\pi^2}~\sum_{\gamma}\bigg|h_k^{(\gamma)}(t)\bigg|^2 \nonumber\\
 &=&\frac{k^2}{\pi^2}\left(\frac{\kappa}{a_\mathrm{b}}\right)^2\left|\frac{H_\mathrm{\omega}\left[\frac{1}{\sqrt{2}} + \sqrt{2}p(t-t_b)\right]}
 {H_\mathrm{\omega}\left[1/\sqrt{2}\right]}\right|^2~~.
 \label{ten power spectrum-g}
\end{eqnarray}
Here we consider the contribution from both the polarization modes of the tensor perturbation. Using Eq.~(\ref{hc-global1}), the tensor power spectrum 
at the horizon crossing comes with the following form,
\begin{eqnarray}
 P_{h}(k,t_h)=\frac{k^2}{\pi^2}\left(\frac{\kappa}{a_\mathrm{b}}\right)^2\left|\frac{H_\mathrm{\omega}\left[\frac{1}{\sqrt{2}} + \sqrt{2}p
 \left(\frac{k}{a_\mathrm{b}\dot{H}_\mathrm{b}}\right)\right]}
 {H_\mathrm{\omega}\left[1/\sqrt{2}\right]}\right|^2~~.
 \label{ten power spectrum-hc-g}
\end{eqnarray}
Therefore the tensor power spectrum is not scale invariant due to the term containing $k/(a_\mathrm{b}\dot{H}_\mathrm{b})$ as well as due to 
$\omega$ (the order of the Hermite polynomial, see Eq.~(\ref{omega})).\\

We now calculate the scalar spectral tilt and the tensor-to-scalar ratio; the scalar tilt is defined in Eq.~(\ref{sp18-g}), 
while the tensor-to-scalar ratio is given by,
\begin{eqnarray}
 r = \frac{P_h}{P_{\Psi}}\bigg|_{h}~~,\label{r}
\end{eqnarray}
where the suffix 'h' denotes the horizon crossing instant. Clearly the $n_s$ and $r$ depends on the parameter $t_s$, i.e the instant when the Type-IV 
singularity occurs. As we have mentioned earlier that depending on whether $t_s<0$ or $t_s>0$ or $t_s = 0$, the Type-IV singularity appears before 
the bounce or after the bounce or at the bounce, respectively. Therefore in the following, we will estimate $n_s$ and $r$ separately for these three cases.

\begin{itemize}
 \item \textbf{\underline{For $t_s < 0$}}: In this case, we consider $\frac{t_s}{t_0} = -1$ (for other positive values of $t_s$, the main arguments 
 will not change). The theoretical predictions for $n_s$ and $r$, with respect to the parameter $n$, are given in Table-[\ref{Table-1}] which 
 clearly demonstrates that 
 the scalar power spectrum is highly red tilted and the tensor-to-scalar ratio gets a large value in respect to the 
 Planck results. 
 
   \begin{table}[h]
  \centering
 {%
  \begin{tabular}{|c|c|c|}
   \hline 
    $n$ & Scalar tilt ($n_s$) & Tensor-to-scalar ratio ($r$)\\
   \hline
   0.25 & 0.27 & 13 \\
   \hline
   0.30 & 0.23 & 9  \\
   \hline
   0.40 & 0.22 & 5  \\
   \hline
   0.50 & 0.245 & 3  \\
   \hline
   \hline
  \end{tabular}%
 }
  \caption{Values of $n_s$ and $r$ with the parameter $n$ for $t_s < 0$.}
  \label{Table-1}
 \end{table}
 
 \item \textbf{\underline{For $t_s > 0$}}: Here the Type-IV singularity occurs after the bounce happens and we safely consider 
 $\frac{t_s}{t_0} = 1$. As a result, the $n_s$ and $r$ are predicted, and they are shown in Table-[\ref{Table-2}].
 
 \begin{table}[h]
  \centering
 {%
  \begin{tabular}{|c|c|c|}
   \hline 
    $n$ & Scalar tilt ($n_s$) & Tensor-to-scalar ratio ($r$)\\
   \hline
   0.25 & 0.27 & 13 \\
   \hline
   0.30 & 0.23 & 9  \\
   \hline
   0.40 & 0.22 & 5  \\
   \hline
   0.50 & 0.24 & 3  \\
   \hline
   \hline
  \end{tabular}%
 }
  \caption{Values of $n_s$ and $r$ with the parameter $n$ for $t_s > 0$.}
  \label{Table-2}
 \end{table}
 
 \item \textbf{\underline{For $t_s = 0$}}: Here, the theoretical estimations for $n_s$ and $r$ (with respect to $n$) are shown in Table-[\ref{Table-3}].
 Planck results. 
 
  \begin{table}[h]
  \centering
 {%
  \begin{tabular}{|c|c|c|}
   \hline 
    $n$ & Scalar tilt ($n_s$) & Tensor-to-scalar ratio ($r$)\\
   \hline
   0.25 & 0.19 & 19 \\
   \hline
   0.30 & 0.15 & 15  \\
   \hline
   0.40 & 0.13 & 10  \\
   \hline
   0.50 & 0.135 & 7  \\
   \hline
   \hline
  \end{tabular}%
 }
  \caption{Values of $n_s$ and $r$ with the parameter $n$ for $t_s = 0$.}
  \label{Table-3}
 \end{table}

\end{itemize}

Therefore in all the three cases, the scalar power spectrum is found to be highly red tilted, and moreover, the model predicts a large value of the 
tensor-to-scalar ratio that lies outside of the Planck data, which indicates that the model is not viable with the observational data. 
However as we observe in \cite{Elizalde:2020zcb} that 
the $f(R,\mathcal{G})$ bounce without any finite time singularity, 
where the scalar factor is described by $a(t) = \left(1+a_0(t/t_0)^2\right)^{n}$, 
indeed leads to the simultaneous compatibility of $n_s$ and $r$ with the Planck data. Therefore we may argue that the 
occurrence of the Type-IV singularity considerably affects the bouncing dynamics in the present context, 
which in turn results to the non-viability of the model. 
Actually the appearance of the Type-IV singularity ``globally'' affects the dynamics of the universe compared to the 
case when the singularity is absent. The term ``global'' means that although the singularity occurs at a finite time $t = t_s$, 
it controls the asymptotic evolution of the comoving 
Hubble radius, in particular, the comoving Hubble radius asymptotically goes to zero due to the presence of the singularity. 
In effect of which, the perturbation modes generate near the bounce, 
unlike to the scenario when the singularity is absent and the perturbation modes 
generate far before the bounce in the deep contracting phase. Such generation era of the perturbation modes near the bounce is the main reason 
that why the scalar power spectrum gets red tilted and the tensor-to-scalar ratio has a large value in the present bounce scenario.\\

Thus as a whole, the following arguments can be drawn for the bounce that appears with a Type-IV singularity -- 
(1) if the singularity appears at $t = t_s$, then depending on whether $t_s < 0$ or $t_s > 0$ or $t_s = 0$, the singularity shows before the 
bounce or after the bounce or at the instant of the bounce respectively. (2) In all these three cases, the scalar power spectrum 
gets red tilted and the tensor-to-scalar ratio is too large to be consistent with the Planck data. As we just have mentioned that such inconsistency 
of the observable quantities is due to the occurrence of the Type-IV singularity, in particular, due to the ``global'' effects of the singularity 
on the evolution of the universe.\\

\section{Realization of a bounce with a Type-IV singularity that locally affects the spacetime}\label{sec-local}

\begin{figure}[!h]
\begin{center}
\centering
\includegraphics[width=3.2in,height=2.5in]{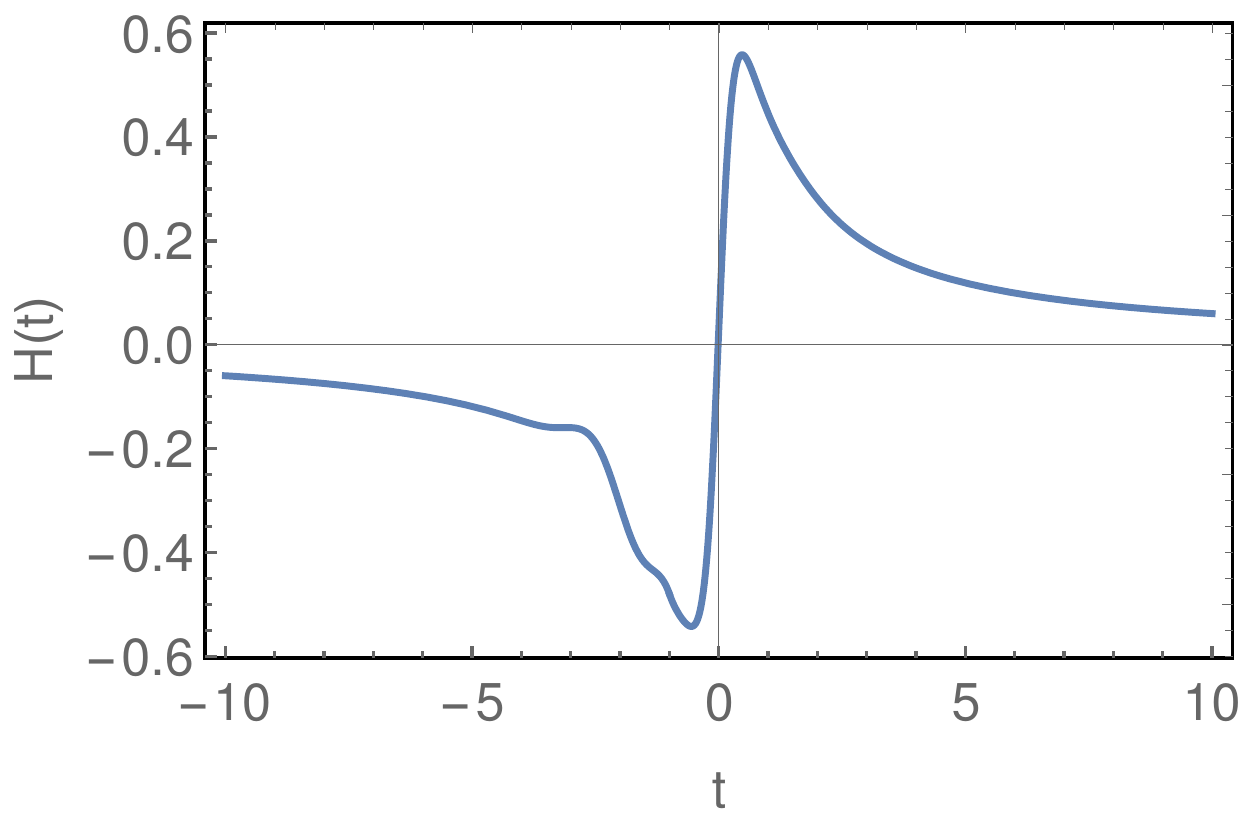}
\includegraphics[width=3.2in,height=2.5in]{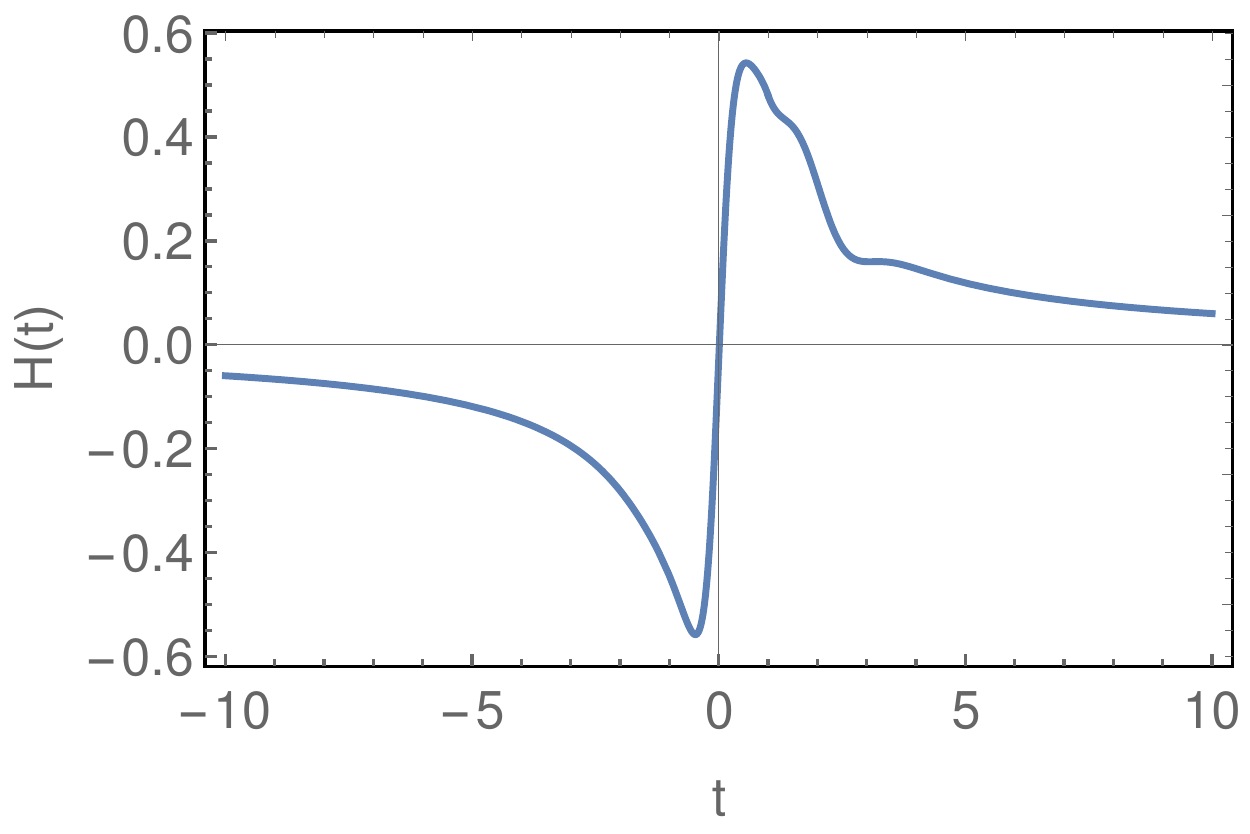}
\caption{$H(t)$ vs. $t$ from Eq.~(\ref{Hubble parameter}). Here we take $n = 0.3$, $a_0 = 4$, $\alpha = 5/3$ and $f_0 = 1$. 
The left and right plots correspond to $t_s = -1\mathrm{By}$ and $t_s  = 1\mathrm{By}$ 
respectively. Moreover $t_0$ is taken to be $1\mathrm{By}$ to make all the time coordinates in the unit 
of Billion year (By). Such values of the parameters lead to the consistency of the observable quantities with the Planck data, 
in the case when the Type-IV singularity locally affects the spacetime, see Fig.[\ref{plot-observable}].}
\label{plot-Hubble}
\end{center}
\end{figure}

In the previous section we have demonstrated that in the case when the Type-IV singularity ``globally'' affects the spacetime, the perturbation modes generate 
near the bounce, and as a result, the observable quantities do not lie within the Planck constraints. 
Based on these findings, it becomes important to examine a bouncing scenario where the Type-IV singularity affects the universe's evolution ``locally'' 
around the time when it occurs. This is the subject of the present section. 
To induce the local effects of the singularity, we introduce a regulating Gaussian factor within the expression of $a_2(t)$ in Eq.(\ref{scale factor-g}). 
Such regulating factor actually controls when the singularity becomes effective. In particular, the scale factor we consider is given by,
\begin{eqnarray}
 a(t) = a_1(t)\times a_2(t) = \left(1+a_0\left(\frac{t}{t_0}\right)^2\right)^{n}\times 
 \exp{\left[\frac{f_0}{(\alpha+1)}\left(\frac{t-t_s}{t_0}\right)^{\alpha+1}e^{-(t-t_s)^2/t_0^2}\right]}~~,
 \label{scale factor}
\end{eqnarray}
where $e^{-(t-t_s)^2}$ acts as the regulating factor, which is peaked around $t=t_s$ i.e at the time when the singularity occurs. The above expression 
is similar to the previous form of the scale factor (see Eq.(\ref{scale factor-g})) except the presence of the regulating factor. 
Once again, the scale factor is written as a product of $a_1(t)$ and $a_2(t)$, where $a_1(t)$ triggers a bounce scenario and $a_2(t)$ 
ensures the occurrence of a finite time singularity at $t = t_s$. Despite the presence of $a_2(t)$, the whole scale factor, i.e $a(t)$, 
predicts a bouncing universe near $t = 0$ -- therefore the presence of $a_2(t)$ results 
to a finite time singularity without jeopardizing the bouncing behaviour of the universe. Moreover the term $e^{-(t-t_s)^2}$ sitting in the expression 
of $a_2(t)$ clearly indicates that $a_2(t)$ becomes effective only around $t = t_s$, otherwise $a_2(t) \approx 1$ away from $t = t_s$ 
and then the universe's evolution is controlled entirely by $a_1(t)$. 
As a result, we may argue that the finite time singularity locally affects the spacetime around the time when it occurs. 
This realizes the importance of the regulating factor to produce a $local$ effect of the 
finite time singularity on the bouncing dynamics, which in turn reflects the significance of the scale factor considered in 
Eq.(\ref{scale factor}) for our present interest. This will be 
clear further from the expression of Hubble parameter defined by $H = \dot{a}/a$. 
Eq.~(\ref{scale factor}) immediately leads to the Hubble parameter as,
\begin{eqnarray}
 H(t) = \frac{1}{t_0}\left[\frac{2a_0n(t/t_0)}{\left(1 + a_0(t/t_0)^2\right)} + f_0\left(\frac{t-t_s}{t_0}\right)^{\alpha}e^{-(t-t_s)^2/t_0^2}
 \left\{1 - \frac{2}{(\alpha+1)}\left(\frac{t-t_s}{t_0}\right)^2\right\}\right]~~.
 \label{Hubble parameter}
\end{eqnarray}
The above expression of $H(t)$ refers to a Type-IV singularity for $\alpha > 1$. The appearance of $e^{-(t-t_s)^2}$ in the expression 
of Eq.~(\ref{Hubble parameter}) clearly indicates that the second term in $H(t)$, which is actually responsible for the singularity, affects 
the evolution of the Hubble parameter only around $t = t_s$, i.e the Type-IV singularity locally affects the spacetime around the time when it occurs. 
Therefore the Hubble parameter of Eq.~(\ref{Hubble parameter}) predicts a bounce at $t \approx 0$. Moreover depending on whether 
$t_s < 0$ or $t_s > 0$, the Type-IV singularity occurs before the bounce or after the bounce, respectively. Using Eq.~(\ref{Hubble parameter}), we 
give the plot of $H(t)$ vs. $t$ in Fig.[\ref{plot-Hubble}] where the left and right plots correspond to $t_s < 0$ and $t_s > 0$ respectively. 

Fig.[\ref{plot-Hubble}] demonstrates the following points about the Hubble parameter -- (1) $H(t)$ 
becomes zero and a increasing function with the cosmic time at $t = 0$, which indicates a bounce of the universe at $t=0$. (2) $H(t)$ is not symmetric 
with respect to the bounce point; this is due to the occurrence of the Type-IV singularity in the present cosmological scenario. Actually the term 
containing $f_0$ in the expression of $H(t)$, which is responsible for the singularity, yields the asymmetric nature of the Hubble parameter. 
(3) The symmetric nature of $H(t)$ seems to deviate 
only around $t = t_s$. This is however expected, because, as we have mentioned earlier, the singularity affects the Hubble parameter locally around 
$t = t_s$. From Eq.~(\ref{scale factor}), the factor $a_2(t) \approx 1$ away from $t = t_s$, and then the universe's evolution is controlled by 
the symmetric scale factor $a_1(t)$.\\

The comoving Hubble radius is defined by $r_h = 1/|aH|$, where $a(t)$ and $H(t)$ are shown above. Therefore at the distant past, the comoving 
Hubble radius turns out to be,
\begin{eqnarray}
 \lim_{t\rightarrow -\infty}~r_h \sim \left|t^{1-2n}\right|~~.\nonumber
\end{eqnarray}
Depending on whether $n < 1/2$ or $n > 1/2$, the asymptotic nature of $r_h$ becomes different, which in turn reveals the generation era of primordial 
perturbation modes. In particular, for $n < 1/2$, 
the comoving Hubble radius in the present context diverges to infinity at the distant past, and as a consequence, the primordial 
perturbation modes generate far away from the bounce at the deep contracting phase. This ensures the 
Bunch-Davies vacuum state of the perturbation at $t \rightarrow -\infty$, and as a result, the horizon problem gets resolved. However 
for $n > 1/2$, the comoving Hubble radius asymptotically goes to zero at both sides of the bounce, and hence the perturbation modes generate 
near the bounce when the Hubble radius is infinite in size to contain all the modes within it. In such case, the perturbation modes at the distant past 
lie outside of the Hubble radius, and thus the horizon problem persists for $n > 1/2$. Based on the above arguments, we will consider $n < 1/2$ so that 
the perturbation modes are within the sub-Hubble regime at the distant past and the horizon problem can be resolved.

%%%%%%%%%%%
%%%%%%%%%%%
%%%%%%%%%%%
%%%%%%%%%%%
%%%%%%%%%%%
%%%%%%%%%%%
%%%%%%%%%%%
%%%%%%%%%%%

\subsection*{Cosmological perturbation and phenomenology of the bounce}\label{sec-perturbation}
As mentioned in the previous section, we take $n < 1/2$, which leads to the generation era of the perturbation modes far before the bounce 
in the sub-Hubble regime. 
Therefore the useful quantities are the background scale factor, the Hubble parameter as well as its derivative (with respect to the cosmic time) and the 
Ricci scalar, during the contracting era. They are given by,
\begin{eqnarray}
 a(t)&=&a_0^{n}\left(\frac{t}{t_0}\right)^{2n}\exp{\left[\frac{f_0}{(\alpha+1)}t^{\alpha + 1}e^{-t^2/t_0^2}\right]}~~~~,~~~~
 H(t) = \frac{2n}{t}\left[1 - \frac{f_0\left(t/t_0\right)^{\alpha+3}e^{-t^2/t_0^2}}{n(\alpha+1)}\right]~~,\nonumber\\
 \dot{H}(t)&=&-\frac{2n}{t^2}\left[1 - \frac{2f_0\left(t/t_0\right)^{\alpha+5}e^{-t^2/t_0^2}}{n(\alpha+1)}\right]~~~~,~~~~
 R(t) = \frac{12n(1-4n)}{t^2}~~,
 \label{useful quantities}
\end{eqnarray}
respectively. Using the expression of $R = R(t)$, we can write the scale factor, the Hubble parameter and its derivative in terms of the Ricci scalar as,
\begin{eqnarray}
a(R)&=&\frac{a_0^{n}}{\left(\widetilde{R}/R_0\right)^{n}}
\left\{1 + \frac{f_0~\exp{\left(-R_0/\widetilde{R}\right)}}{(\alpha+1)\left(\widetilde{R}/R_0\right)^{\frac{\alpha}{2} + \frac{1}{2}}}\right\}~~~~,~~~~
H(R) = -2n\widetilde{R}^{1/2}
\left\{1 - \frac{f_0~\exp{\left(-R_0/\widetilde{R}\right)}}{n(\alpha+1)\left(\widetilde{R}/R_0\right)^{\frac{\alpha}{2} + \frac{3}{2}}}\right\}~,\nonumber\\
\dot{H}(R)&=&-2n\widetilde{R}
\left\{1 - \frac{2f_0~\exp{\left(-R_0/\widetilde{R}\right)}}{n(\alpha+1)\left(\widetilde{R}/R_0\right)^{\frac{\alpha}{2} + \frac{5}{2}}}\right\}~~~~,~~~~
\dot{h}(R) = \frac{h_0(2n+1)}{\widetilde{R}^n}
\left\{1 + \frac{f_0~\exp{\left(-R_0/\widetilde{R}\right)}}{(\alpha+1)\left(\widetilde{R}/R_0\right)^{\frac{\alpha}{2} + \frac{1}{2}}}\right\}~,
\label{quantities-deep-contracting}
\end{eqnarray}
where in the last equality, we write $\dot{h} = \dot{h}(R)$ from Eq.~(\ref{constraint on coupling}), with $h_0$ is a constant having mass dimension 
[1+2n]. Moreover $R_0 = \frac{1}{t_0^2}$ and $\widetilde{R}(t) = \frac{R(t)}{12n(1-4n)}$ in the above expressions. 
Using $H(R)$ and $\dot{h}(R)$ from Eq.~(\ref{quantities-deep-contracting}), we determine the functions $Q_i$ in the context of 
the ghost free Gauss-Bonnet theory of gravity \cite{Hwang:2005hb,Noh:2001ia,Hwang:2002fp} as,
\begin{align}
Q_a=&-8\dot{h}H^2 = -32h_0n^2(1+2n)\widetilde{R}^{1-n}
\left\{1 - \frac{2f_0~\exp{\left(-R_0/\widetilde{R}\right)}}{n(\alpha+1)\left(\widetilde{R}/R_0\right)^{\frac{\alpha}{2} + \frac{3}{2}}}\right\}\, , \nonumber\\
Q_b=&-16\dot{h}H = 32h_0n(1+2n)\widetilde{R}^{\frac{1}{2} - n}
\left\{1 - \frac{f_0~\exp{\left(-R_0/\widetilde{R}\right)}}{n(\alpha+1)\left(\widetilde{R}/R_0\right)^{\frac{\alpha}{2} + \frac{3}{2}}}\right\}\, ,\nonumber\\
Q_c=&Q_d = 0 \, ,\nonumber\\
Q_e=&-32\dot{h}\dot{H} = 64h_0n(1+2n)\widetilde{R}^{1-n}
\left\{1 - \frac{2f_0~\exp{\left(-R_0/\widetilde{R}\right)}}{n(\alpha+1)\left(\widetilde{R}/R_0\right)^{\frac{\alpha}{2} + \frac{5}{2}}}\right\}\, ,\nonumber\\
Q_f=&16 \left[ \ddot{h} - \dot{h}H \right] = 0 \, ,
\label{Q-s}
\end{align}
respectively, we will use these expressions frequently in the perturbation calculations. 
Recall, $h_0$ has mass dimension [1+2n] and thus from dimensional analysis, we can consider,
\begin{eqnarray}
\kappa^2h_0 = \left(t_0\right)^{1-2n} = \left(\frac{1}{R_0}\right)^{\frac{1}{2} - n}\, ,
\label{h0-dimension}
\end{eqnarray}
where $R_0$ is a positive constant. The parameters $h_0$ and $t_0$ are considered to be related by the above fashion, and $R_0$ can be regarded to be 
the replacement of both of them. Consequently Eq.~(\ref{final form2}) 
along with the Eq.~(\ref{quantities-deep-contracting}) immediately lead to the Lagrange multiplier function as,
\begin{eqnarray}
\mu^4\lambda = -\frac{4n\widetilde{R}}{\kappa^2}\left[1 - 16n(1+2n)\left(\frac{\widetilde{R}(t)}{R_0}\right)^{\frac{1}{2} - n}
\left\{1 - \frac{f_0~\exp{\left(-R_0/\widetilde{R}\right)}}{n(\alpha+1)\left(\widetilde{R}/R_0\right)^{\frac{\alpha}{2} + \frac{3}{2}}}\right\}\right]\, .
\label{LM-deep-contracting}
\end{eqnarray}
We will use the these expressions in addressing the evolution of scalar and tensor perturbations in the following two subsections, respectively. 

\subsubsection*{Scalar perturbation}
As the perturbation modes generate at the distant past, it will be useful to work in the comoving gauge, in which case, 
the second order perturbed action for curvature perturbation (symbolized by $\Psi(t,\vec{x})$) 
is given by \cite{Hwang:2005hb,Noh:2001ia,Hwang:2002fp},
\begin{align}
\delta S_{\psi} = \int dt d^3\vec{x} a(t) z(t)^2\left[\dot{\Psi}^2
 - \frac{c_s^2}{a^2}\left(\partial_i\Psi\right)^2\right]\, .
\label{sp2}
\end{align}
Here in the context of Lagrange multiplier $f(R,\mathcal{G})$ gravity, $z(t)$ and $c_s^2$ have the following forms \cite{Hwang:2005hb},
\begin{align}
z(t) = \frac{a(t)}{H + \frac{Q_a}{2F + Q_b}} \sqrt{-\mu^4\lambda + \frac{3Q_a^2 + Q_aQ_e}{2F + Q_b}}
\label{sp3}
\end{align}
and
\begin{align}
c_{s}^{2} = 1 + \frac{Q_aQ_e/\left(2F + Q_b\right)}{-\mu^4\lambda + 3\frac{Q_a^2}{2F + Q_b}} \, ,
\label{speed scalar perturbation}
\end{align}
respectively, with the functions $Q_i$ being defined earlier and $F=\frac{1}{2\kappa^2}$. 
From Eq.~(\ref{sp2}), it is clear that the kinetic term of the curvature perturbation comes with a positive sign under the condition $z^2(t) > 0$. Later, 
we will show that this condition, i.e $z^2(t)>0$, indeed holds in the present context, which in turn ensures the stability 
of the curvature perturbation. 
By using Eq.~(\ref{quantities-deep-contracting}) and Eq.~(\ref{Q-s}), 
we determine various terms present in the expression of $z(t)$ as follows,
\begin{align}
\frac{a(t)}{H + \frac{Q_a}{2F + Q_b}} = -\frac{a_0^{n}}{\sqrt{\widetilde{R}}}\left(\frac{R_0}{\widetilde{R}}\right)^{n}
\left[2n\left\{1 + 16n(1+2n)\left(\frac{\widetilde{R}}{R_0}\right)^{\frac{1}{2} - n}
\left\{1 - \frac{2f_0~\exp{\left(-R_0/\widetilde{R}\right)}}{n(\alpha+1)\left(\widetilde{R}/R_0\right)^{\frac{\alpha}{2} + \frac{3}{2}}}\right\}
+ \mathcal{O}\left(\frac{\widetilde{R}}{R_0}\right)^{1-2n}\right\}\right]^{-1}
\nonumber
\end{align}
and
\begin{eqnarray}
&-&\mu^4\lambda + \frac{3Q_a^2}{2F + Q_b} + \frac{Q_aQ_e}{2F + Q_b}\nonumber\\
&=&\frac{4n\widetilde{R}}{\kappa^2}\left[1 - 16n(1+2n)\left(\frac{\widetilde{R}}{R_0}\right)^{\frac{1}{2} - n}
\left\{1 - \frac{f_0~\exp{\left(-R_0/\widetilde{R}\right)}}{n(\alpha+1)\left(\widetilde{R}/R_0\right)^{\frac{\alpha}{2} + \frac{3}{2}}}\right\}
+ \mathcal{O}\left(\frac{\widetilde{R}}{R_0}\right)^{1-2n}\right]\, ,
\nonumber
\end{eqnarray}
respectively. Consequently the form of $z(t)$ from Eq.~(\ref{sp3}) becomes,
\begin{align}
z(t) = -\frac{a_0^n}{\kappa \left(\widetilde{R}/R_0\right)^{n}}~\frac{\sqrt{P(R)}}{Q(R)}
\label{sp4}
\end{align}
where $P(R)$ and $Q(R)$ have the following forms,
\begin{align}
P(R) = 4n\left[1 - 16n(1+2n)\left(\frac{\widetilde{R}}{R_0}\right)^{\frac{1}{2} - n}
\left\{1 - \frac{f_0~\exp{\left(-R_0/\widetilde{R}\right)}}{n(\alpha+1)\left(\widetilde{R}/R_0\right)^{\frac{\alpha}{2} + \frac{3}{2}}}\right\}
+ \mathcal{O}\left(\frac{\widetilde{R}}{R_0}\right)^{1-2n}\right]\, ,
\label{P}
\end{align}
and
\begin{align}
Q(R) = 2n\left[1 + 16n(1+2n)\left(\frac{\widetilde{R}}{R_0}\right)^{\frac{1}{2} - n}
\left\{1 - \frac{2f_0~\exp{\left(-R_0/\widetilde{R}\right)}}{n(\alpha+1)\left(\widetilde{R}/R_0\right)^{\frac{\alpha}{2} + \frac{3}{2}}}\right\}
+ \mathcal{O}\left(\frac{\widetilde{R}}{R_0}\right)^{1-2n}\right]\, ,
\label{Q}
\end{align}
respectively. Note the dependence of $z(t)$ on the parameter $f_0$ which actually arises due to the occurrence of the Type-IV singularity during the 
universe's evolution. Our intention is to examine how the observable quantities (like the scalar spectral index and the tensor-to-scalar ratio) 
depend on $f_0$ when the Type-IV singularity affects the spacetime locally around $t = t_s$. 
As demonstrated earlier, the perturbation modes generate during the late contracting phase when the Ricci scalar satisfies 
the condition like $\frac{\widetilde{R}}{R_0} \ll 1$ as $\widetilde{R} \rightarrow 0$ at $t \rightarrow -\infty$ 
(the numerical estimation of $\widetilde{R}/R_0$ is given after Eq.~(\ref{hc-1})). As a consequence, Eq.~(\ref{sp4}) leads to 
$z^2(t) > 0$ which makes the curvature perturbation stable.

It will be more useful if we transform the time coordinate to conformal time defined by $\eta = \int\frac{dt}{a(t)}$. 
Due to Eq.~(\ref{scale factor}) the scale factor at distant past behaves as $a(t) \sim t^{2n}$, and hence the corresponding conformal time comes as,
\begin{eqnarray}
\eta(t) = \left[\frac{1}{a_0^n(1-2n)}\right]t^{1-2n}\, .
\label{conformal time}
\end{eqnarray}
Recall that $n < \frac{1}{2}$ in order to resolve the horizon issue, due to which, $\eta(t)$ appears to be a monotonic increasing function 
of the cosmic time. With Eq.~(\ref{conformal time}), we obtain the Ricci scalar in terms of $\eta$ as, 
\begin{eqnarray}
\widetilde{R}(\eta) = \frac{1}{\left[a_0^n(1-2n)\right]^{2/(1-2n)}}\times\frac{1}{\eta^{2/(1-2n)}} \propto \frac{1}{\eta^{2/(1-2n)}}\, .
\label{ricci-scalar-conformal-time}
\end{eqnarray}
Using the above expression of $R(\eta)$ into Eq.~(\ref{sp4}), we get $z(\eta)$ as follows,
\begin{eqnarray}
z(\eta) \propto \left(\frac{\sqrt{P(\eta)}}{Q(\eta)}\right)\eta^{2n/(1-2n)}\, ,
\label{z-eta}
\end{eqnarray}
with $P(\eta) = P(R(\eta))$ and $Q(\eta) = Q(R(\eta))$. 
Consequently we determine the factor $\frac{1}{z}\frac{d^2z}{d\eta^2}$ (which is essential for solving the scalar Mukhanov-Sasaki equation),
\begin{eqnarray}
\frac{1}{z}\frac{d^2z}{d\eta^2} = \frac{\xi(\xi - 1)}{\eta^2}\left\{1 + 24\left(1-4n^2\right)\left(\frac{\widetilde{R}}{R_0}\right)^{\frac{1}{2} - n}
\left\{1 - \frac{10f_0~\exp{\left(-R_0/\widetilde{R}\right)}}{3n(\alpha+1)(1-2n)(1-4n)\left(\widetilde{R}/R_0\right)^{\frac{\alpha}{2} + \frac{7}{2}}}\right\}
+ \mathcal{O}\left(\frac{\widetilde{R}}{R_0}\right)^{1-2n}\right\}
\label{derivative-z-eta}
\end{eqnarray}
where $\xi = \frac{2n}{(1-2n)}$ and we use $\frac{d\widetilde{R}}{d\eta} = \frac{-2}{(1-2n)}\frac{\widetilde{R}}{\eta}$. 
Furthermore with the aforementioned expressions of $Q_i$ (see Eq.~(\ref{Q-s})), 
the speed of the scalar perturbation from Eq.~(\ref{speed scalar perturbation}) turns out to be,
\begin{eqnarray}
c_s^2 = 1 + \mathcal{O}\left(\frac{\widetilde{R}}{R_0}\right)^{1-2n}\, .
\label{sound-speed}
\end{eqnarray}
At this stage, we now introduce the scalar Mukhanov-Sasaki (MS) equation for the variable $v(\eta,\vec{x}) = z(\eta)\Psi(\eta,\vec{x})$ 
(also known as MS variable), 
\begin{eqnarray}
\frac{d^2v_k(\eta)}{d\eta^2} + \left(c_s^2k^2 - \frac{1}{z}\frac{d^2z}{d\eta^2}\right)v_k(\eta) = 0\, ,
\label{scalar-MS-equation}
\end{eqnarray}
where $v_k(\eta)$ is the Fourier mode for $v(\eta,\vec{x})$. 
Clearly the evolution of $v_k(\eta)$ depends on the background evolution through the factors $z''/z$ and $c_s^2$ (the overprime denotes the derivative 
with respect to $\eta$). Due to the condition $\frac{\widetilde{R}}{R_0} \ll 1$ (as depicted after Eq.~(\ref{Q})), $z''/z$ and $c_s^2$ 
can be expressed as,
\begin{eqnarray}
\frac{1}{z}\frac{d^2z}{d\eta^2}&=&\frac{\xi(\xi - 1)}{\eta^2}\left[1 + 24\left(1-4n^2\right)\left(\frac{\widetilde{R}}{R_0}\right)^{\frac{1}{2} - n}
\left\{1 - \frac{10f_0~\exp{\left(-R_0/\widetilde{R}\right)}}{3n(\alpha+1)(1-2n)(1-4n)\left(\widetilde{R}/R_0\right)
^{\frac{\alpha}{2} + \frac{7}{2}}}\right\}
\right]~~,\nonumber\\
c_s^2&=&1\, ,
\label{approximate-behaviour}
\end{eqnarray}
respectively, where we retain the terms up-to the order $\left(\widetilde{R}/R_0\right)^{\frac{1}{2} - n}$. 
Furthermore $\frac{\widetilde{R}}{R_0} \ll 1$ along with $n < 1/2$ (in order to generate the perturbation modes far before the bounce and 
consequently to resolve the horizon issue) clearly depict that 
the term $\left(\widetilde{R}/R_0\right)^{\frac{1}{2} - n}$ within the paranthesis can be safely considered to be small during the contracting era. 
As a result, $z''/z$ becomes proportional to $1/\eta^2$, i.e., $\frac{1}{z}\frac{d^2z}{d\eta^2} = \sigma/\eta^2$, with,
\begin{eqnarray}
\sigma = \xi(\xi - 1)\left[1 + 24\left(1-4n^2\right)\left(\frac{\widetilde{R}}{R_0}\right)^{\frac{1}{2} - n}
\left\{1 - \frac{10f_0~\exp{\left(-R_0/\widetilde{R}\right)}}{3n(\alpha+1)(1-2n)(1-4n)\left(\widetilde{R}/R_0\right)
^{\frac{\alpha}{2} + \frac{7}{2}}}\right\}
\right]\, ,
\label{sigma}
\end{eqnarray}
which is approximately a constant during the era when the perturbation modes generate deep inside the Hubble radius. Accordingly, along with 
$c_s^2 = 1$, we solve $v_k(\eta)$ from Eq.~(\ref{scalar-MS-equation}) and is given by,
\begin{eqnarray}
v(k,\eta) = \frac{\sqrt{\pi|\eta|}}{2}\left[c_1(k)H_{\nu}^{(1)}(k|\eta|) + c_2(k)H_{\nu}^{(2)}(k|\eta|)\right]\, ,
\label{scalar-MS-solution}
\end{eqnarray}
with $\nu = \sqrt{\sigma + \frac{1}{4}}$, and moreover, 
$H_{\nu}^{(1)}(k|\eta|)$ and $H_{\nu}^{(2)}(k|\eta|)$ are the Hermite functions (having order 
$\nu$) of first and second kind, respectively. Here $c_1$, $c_2$ are integration constants which can be determined from the initial condition of 
the MS variable. The Bunch-Davies vacuum state is considered to be the initial state for $v_k(\eta)$, 
in particular, $\lim_{k|\eta| \gg 1}v(k,\eta) = \frac{1}{\sqrt{2k}}e^{-ik\eta}$. The Bunch-Davies initial condition is ensured from the fact 
that the perturbation modes generate in the deep sub-Hubble regime (when all the perturbation modes lie within the Hubble radius). 
The Bunch-Davies condition immediately leads to $c_1 = 0$ and $c_2 = 1$, respectively. 
Consequently, the scalar power spectrum for $k$th mode turns out to be,
\begin{eqnarray}
\mathcal{P}_{\Psi}(k,\eta) = \frac{k^3}{2\pi^2} \left| \frac{v(k,\eta)}{z(\eta)} \right|^2 
= \frac{k^3}{2\pi^2} \left| \frac{\sqrt{\pi|\eta|}}{2z (\eta)}H_{\nu}^{(2)}(k|\eta|) \right|^2\, ,
\label{scalar-power-spectrum}
\end{eqnarray}
where in the second equality, we use the solution of $v(k,\eta)$. 
The $k$-th mode satisfies the relation $k = |aH|$ at the instant of horizon crossing, which, by using 
Eq.~(\ref{quantities-deep-contracting}), is obtained as,
\begin{eqnarray}
k = \frac{1}{\left| \eta_h\right|}\left(\frac{2n}{1-2n}\right) \quad \Rightarrow \quad k\left| \eta_h\right| = \frac{2n}{1-2n}\, ,
\label{hc-1}
\end{eqnarray}
where the suffix 'h' represents the horizon crossing instant. Eq.~(\ref{hc-1}) estimates the horizon crossing time for large scale modes, 
in particular for $k = 0.05\mathrm{Mpc}^{-1}$ (around which we will determine he observable quantities), as 
$\eta_h \approx -13\,\mathrm{By}$. This is however expected, because the large scale modes re-enter the horizon around the present epoch when the age of the 
universe is nearly $\approx 13.5\mathrm{By}$. Since the universe is almost symmetric with respect to the bounce point (except around $t = t_s$ when the 
Type-IV singularity occurs), one can already guess that the large scale modes cross the horizon during the contracting phase nearly at 
$\eta_h \approx -13\mathrm{By}$, which is also reflected from Eq.~(\ref{hc-1}). Consequently we estimate the Ricci scalar at the horizon crossing of the large 
scale modes, in particular, we get $\frac{\widetilde{R}}{R_0} \sim 10^{-6}$ (where we consider $n = 0.3$, $R_0 = 1\mathrm{By}^{-2}$ and 
$a_0 \sim \mathcal{O}(1)$: we will show that such considerations of $n$, $R_0$ and $a_0$ indeed are consistent 
with the viability of the observable quantities in respect to the Planck data). This justifies the condition $\frac{\widetilde{R}}{R_0} \ll 1$ 
which we have considered earlier in the expression of $z(t)$.

Eq.~(\ref{hc-1}) depicts the sub-Hubble and super-Hubble regime of $k$-th mode as,
\begin{align}
k\left|\eta\right|>\frac{2n}{1-2n}\, :& \, \mathrm{sub~Hubble~regime}\, ,\nonumber\\
k\left|\eta\right|<\frac{2n}{1-2n}\, :& \, \mathrm{super~Hubble~regime}\, .
\label{hc-2}
\end{align}
As a result, the scalar power spectrum (from Eq.~(\ref{scalar-power-spectrum})) in the super-Hubble regime can be expressed as,
\begin{eqnarray}
\mathcal{P}_{\Psi}(k,\eta) = \left[\left(\frac{1}{2\pi}\right)\frac{1}{z\left|\eta\right|}\frac{\Gamma(\nu)}{\Gamma(3/2)}\right]^2
\left(\frac{k|\eta|}{2}\right)^{3-2\nu}\, ,
\label{scalar-power-spectrum-superhorizon}
\end{eqnarray}
with recall that $\nu = \sqrt{\sigma + \frac{1}{4}}$. 
By using Eq.~(\ref{scalar-power-spectrum-superhorizon}), we can determine the spectral tilt of the primordial curvature perturbations (symbolized 
by $n_s$). Clearly $\nu$ depends on $f_0$, and thus the occurrence of the Type-IV 
singularity affects the scalar power spectrum as well as the corresponding spectral tilt. 
However before proceeding to calculate $n_s$, we will determine the tensor power spectrum, which is necessary for 
the prediction of the tensor-to-scalar ratio.

\subsubsection*{Tensor perturbation}
The tensor perturbation over FRW metric is,
\begin{align}
ds^2 = -dt^2 + a(t)^2\left(\delta_{ij} + h_{ij}\right)dx^idx^j\, ,
\label{tp1}
\end{align}
where $h_{ij}(t,\vec{x})$ is the tensor perturbation variable, and the corresponding 
tensor perturbed action (up-to quadratic order) is given by \cite{Hwang:2005hb,Noh:2001ia,Hwang:2002fp,Kawai:1999pw,Kawai:1998ab},
\begin{align}
\delta S_{h} = \int dt d^3\vec{x} a(t) z_T(t)^2\left[\dot{h}_{ij}\dot{h}^{ij}
 - \frac{1}{a^2}\left(\partial_kh_{ij}\right)^2\right]\, .
\label{tp2}
\end{align}
In the Lagrange multiplier Gauss-Bonnet gravity theory, the function $z_T$ is \cite{Hwang:2005hb},
\begin{align}
z_T(t) = a\sqrt{F + \frac{1}{2}Q_b}\, ,
\label{tp3}
\end{align}
where $F = \frac{1}{2\kappa^2}$ and the $Q_b$ is given in Eq.~(\ref{Q-s}). 
Eq.~(\ref{tp2}) indicates that the speed of the tensor perturbation (or equivalently the gravitational waves) is equal 
to unity -- this is due to the fact that the Gauss-Bonnet coupling in the present context satisfies $\ddot{h} = \dot{h}H$ which 
in turn makes $c_T^2 = 1$, and consequently, the model gets compatible with the GW170817 event. 
The scale factor from Eq.~(\ref{quantities-deep-contracting}) immediately leads to the following expression of $z_T$ as, 
\begin{align}
z_T=\frac{a_0^n}{\sqrt{2}\kappa\widetilde{R}^n}\left[1 + 16n(1+2n)\left(\frac{\widetilde{R}}{R_0}\right)^{\frac{1}{2} - n}
\left\{1 - \frac{f_0~\exp{\left(-R_0/\widetilde{R}\right)}}{n(\alpha+1)\left(\widetilde{R}/R_0\right)^{\frac{\alpha}{2} + \frac{3}{2}}}\right\}\right]\, ,
\label{zT-1}
\end{align}
Therefore $z_T^2$ is positive, which results to the stability of the tensor perturbation. 
By using Eq.~(\ref{ricci-scalar-conformal-time}), i.e $\widetilde{R}(\eta) \propto \eta^{-2/(1-2n)}$, we determine $z_T$ in terms of the conformal time as,
\begin{eqnarray}
z_T(\eta) \propto S(R(\eta))\eta^{2n/(1-2n)}\, ,
\label{zT-2}
\end{eqnarray}
with $S(R(\eta))$ is given by,
\begin{eqnarray}
S(R(\eta)) = 1 + 16n(1+2n)\left(\frac{\widetilde{R}}{R_0}\right)^{\frac{1}{2} - n}
\left\{1 - \frac{f_0~\exp{\left(-R_0/\widetilde{R}\right)}}{n(\alpha+1)\left(\widetilde{R}/R_0\right)^{\frac{\alpha}{2} + \frac{3}{2}}}\right\}\, .
\label{S}
\end{eqnarray}
Accordingly we calculate $z_T''/z_T$,
\begin{eqnarray}
\frac{1}{z_T}\frac{d^2z_T}{d\eta^2} = \frac{\xi(\xi-1)}{\eta^2}\left[1 - 16(1-4n^2)\left(\frac{\widetilde{R}}{R_0}\right)^{\frac{1}{2} - n}
\left\{1 - \frac{2f_0~\exp{\left(-R_0/\widetilde{R}\right)}}{n(\alpha+1)(1-2n)(1-4n)\left(\widetilde{R}/R_0\right)^{\frac{\alpha}{2} + \frac{7}{2}}}\right\}
\right]\, ,
\label{derivative-zT}
\end{eqnarray}
where recall that $\xi = \frac{2n}{1-2n}$, and we use $\frac{d\widetilde{R}}{d\eta} = \frac{-2}{(1-2n)}\frac{\widetilde{R}}{\eta}$ to arrive at the above 
expression. 
The above expression will be useful for solving the tensor Mukhanov-Sasaki equation. 
Due to the condition $\frac{\widetilde{R}}{R_0} \ll 1$ (as demonstrated earlier) along with $n < 1/2$, the tern containing 
$\left(\frac{\widetilde{R}}{R_0}\right)^{\frac{1}{2} - n}$ can be safely regarded to be small during the contracting phase. As a result, 
$z_T''/z_T$ gets proportional to $1/\eta^2$, i.e., $\frac{1}{z_T}\frac{d^2z_T}{d\eta^2} = \sigma_T/\eta^2$, with
\begin{eqnarray}
\sigma_T = \xi(\xi - 1)\left[1 - 16(1-4n^2)\left(\frac{\widetilde{R}}{R_0}\right)^{\frac{1}{2} - n}
\left\{1 - \frac{2f_0~\exp{\left(-R_0/\widetilde{R}\right)}}{n(\alpha+1)(1-2n)(1-4n)\left(\widetilde{R}/R_0\right)^{\frac{\alpha}{2} + \frac{7}{2}}}\right\}
\right]\, .
\label{sigma-T}
\end{eqnarray}
In effect, the tensor Mukhanov-Sasaki (MS) equation becomes,
\begin{align}
\frac{d^2v_T(k,\eta)}{d\eta^2} + \left(k^2 - \frac{\sigma_T}{\eta^2}\right)v_T(k,\eta) = 0\, ,
\label{tensor-MS-equation}
\end{align}
with $v_T(k,\eta)$ is the Fourier transformed quantity of the tensor MS variable which is defined by $\left(v_T\right)_{ij} = z_Th_{ij}$. 
Considering the Bunch-Davies initial condition for $v_T(k,\eta)$, i.e $\lim_{k|\eta| \gg 1}v_T(k,\eta) = \frac{1}{\sqrt{2k}}e^{-ik\eta}$, 
we solve Eq.~(\ref{tensor-MS-equation}) and is given by,
\begin{eqnarray}
v_T(k,\eta) = \frac{\sqrt{\pi\left|\eta\right|}}{2} H_{\theta}^{(2)}(k\left|\eta\right|)
\label{tensor-MS-solution}
\end{eqnarray}
with $\theta = \sqrt{\sigma_T + \frac{1}{4}}$ and $H_{\theta}^{(2)}(k\left|\eta\right|)$ represents the Hermite function of second kind having order $\theta$. 
Consequently the tensor power spectrum for $k$-th mode in the superhorizon scale (when the relevant modes are outside of the Hubble radius and 
satisfying $k|\eta| \ll 1$ from Eq.~(\ref{hc-2})) comes with the following expression,
\begin{align}
\mathcal{P}_{T}(k,\tau) = 2\left[\frac{1}{2\pi}\frac{1}{z_T\left|\eta\right|}\frac{\Gamma(\theta)}{\Gamma(3/2)}\right]^2 \left(\frac{k|\eta|}{2}
\right)^{3 - 2\theta}\, ,
\label{tensor-power-spectrum}
\end{align}
where we consider the contributions from both the polarization modes.\\

Having set the stage, we now calculate the observable quantities like the scalar spectral index ($n_s$) and the tensor-to-scalar ratio ($r$) respectively. 
They are defined by,
\begin{eqnarray}
n_s = 1 + \left. \frac{\partial \ln{\mathcal{P}_{\Psi}}}{\partial \ln{k}} \right|_{h} \, , \quad r=\mathcal{P}_T/\mathcal{P}_{\Psi}\, ,
\label{obs-1}
\end{eqnarray}
where the suffix 'h' represents the horizon crossing instant of the large scale modes ($\sim 0.05\mathrm{Mpc}^{-1}$) around which we will estimate 
the observable indices. According to the recent Planck data, $n_s$ and $r$ are constrained by \cite{Akrami:2018odb},
\begin{eqnarray}
n_s = 0.9649 \pm 0.0042 \quad \mbox{and} \quad r < 0.064 \, ,
\label{observable-Planck constraint}
\end{eqnarray}
respectively. 
Due to Eq.~(\ref{scalar-power-spectrum-superhorizon}) and 
Eq.~(\ref{tensor-power-spectrum}), we determine the final forms of $n_s$ and $r$ in the present context as,
\begin{eqnarray}
n_s = 4 - \sqrt{1 + 4\sigma_h} \, , \quad r = 2\left[\frac{z(\eta_h)}{z_T(\eta_h)}\frac{\Gamma(\theta)}{\Gamma(\nu)}\right]^2
\left( k\left|\eta_h\right| \right)^{2(\nu-\theta)}\, ,
\label{obs-2}
\end{eqnarray}
where all the quantities are evaluated at horizon crossing of large scale modes, in particular,
\begin{align}
\nu=&\sqrt{\sigma_h + \frac{1}{4}}\, ; \quad \sigma_h = \xi(\xi - 1)\left[1 + 24\left(1-4n^2\right)\left(\frac{\widetilde{R}_h}{R_0}\right)^{\frac{1}{2} - n}
\left\{1 - \frac{10f_0~\exp{\left(-R_0/\widetilde{R}_h\right)}}{3n(\alpha+1)(1-2n)(1-4n)\left(\widetilde{R}_h/R_0\right)
^{\frac{\alpha}{2} + \frac{7}{2}}}\right\}
\right]\, ,\nonumber\\
\theta=&\sqrt{\sigma_{T,h} + \frac{1}{4}} \, ; \quad \sigma_{T,h} = \xi(\xi - 1)\left[1 - 16(1-4n^2)\left(\frac{\widetilde{R}_h}{R_0}\right)^{\frac{1}{2} - n}
\left\{1 - \frac{2f_0~\exp{\left(-R_0/\widetilde{R}_h\right)}}{n(\alpha+1)(1-2n)(1-4n)\left(\widetilde{R}_h/R_0\right)^{\frac{\alpha}{2} + \frac{7}{2}}}\right\}
\right]\, ,\nonumber\\
z(\eta_h)=&-\frac{1}{\sqrt{n}}\left(\frac{a_0^n}{\kappa\widetilde{R}_h^{n}}\right)\left[1 - 24n(1+2n)
\left(\frac{\widetilde{R}_h}{R_0}\right)^{\frac{1}{2} - n}
\left\{1 - \frac{5f_0~\exp{\left(-R_0/\widetilde{R}_h\right)}}{3n(\alpha+1)\left(\widetilde{R}_h/R_0\right)^{\frac{\alpha}{2} + \frac{3}{2}}}\right\}\right]\, ,\nonumber\\
z_T(\eta_h)=&\frac{1}{\sqrt{2}}\left(\frac{a_0^n}{\kappa\widetilde{R}_h^{n}}\right)\left[1 + 16n(1+2n)
\left(\frac{\widetilde{R}_h}{R_0}\right)^{\frac{1}{2} - n}
\left\{1 - \frac{f_0~\exp{\left(-R_0/\widetilde{R}_h\right)}}{n(\alpha+1)\left(\widetilde{R}_h/R_0\right)^{\frac{\alpha}{2} + \frac{3}{2}}}\right\}\right]\, .
\label{obs-3}
\end{align}
Here we would like to mention that the dependence of $n_s$ and $r$ on the parameter $f_0$ actually 
decodes the possible effects of the Type-IV singularity on the observable quantities. 
Clearly the above expressions contain $\widetilde{R}_h$ which is the Ricci scalar at the horizon crossing of the large scale modes. Hence from 
Eq.~(\ref{ricci-scalar-conformal-time}), one may write,
\begin{eqnarray}
\widetilde{R}_h = \left[\frac{1}{a_0^n(1-2n)\left|\eta_h\right|}\right]^{2/(1-2n)}\, ,
\label{obs-4}
\end{eqnarray}
with $\eta_h$ is shown in Eq.~(\ref{hc-1}), in particular, 
\begin{eqnarray}
\left|\eta_h\right| = \left(\frac{2n}{1-2n}\right)\frac{1}{k} \approx \left(\frac{2n}{1-2n}\right)\times13\,\mathrm{By}\, .
\label{obs-5}
\end{eqnarray}
Here we use $k = 0.05\mathrm{Mpc}^{-1}$ which crosses the horizon during the contracting phase nearly around $\approx -13\mathrm{By}$.  
Plugging back the above expression of $\left|\eta_h\right|$ into Eq.~(\ref{obs-4}), we get $\widetilde{R}_h$ in terms of $n$ and $a_0$:
\begin{eqnarray}
\widetilde{R}_h = \left[\frac{1}{26na_0^n}\right]^{2/(1-2n)}\mathrm{By}^{-2}\, .
\label{obs-6}
\end{eqnarray}
Thus as a whole, the theoretical expressions of $n_s$ and $r$ depend on the parameters $n$, $a_0$ and $f_0$. Here we would like to mention that 
the scalar tilt as well as the tensor-to-scalar ratio do not depend on the parameter $t_s$ (the time when the Type-IV singularity occurs). This is however 
expected, because the singularity affects the spacetime locally around the finite time $t = t_s$ and the perturbation modes generate 
in the deep contracting phase where the singularity provides almost no effects on the universe's evolution. This is unlike to the previous scenario where 
the Type-IV singularity globally affects the spacetime and, as a result, the observable quantities are found to depend on $t_s$, see the discussion 
after the Table-[\ref{Table-3}].

\begin{figure}[!h]
\begin{center}
\centering
\includegraphics[width=3.0in,height=3.0in]{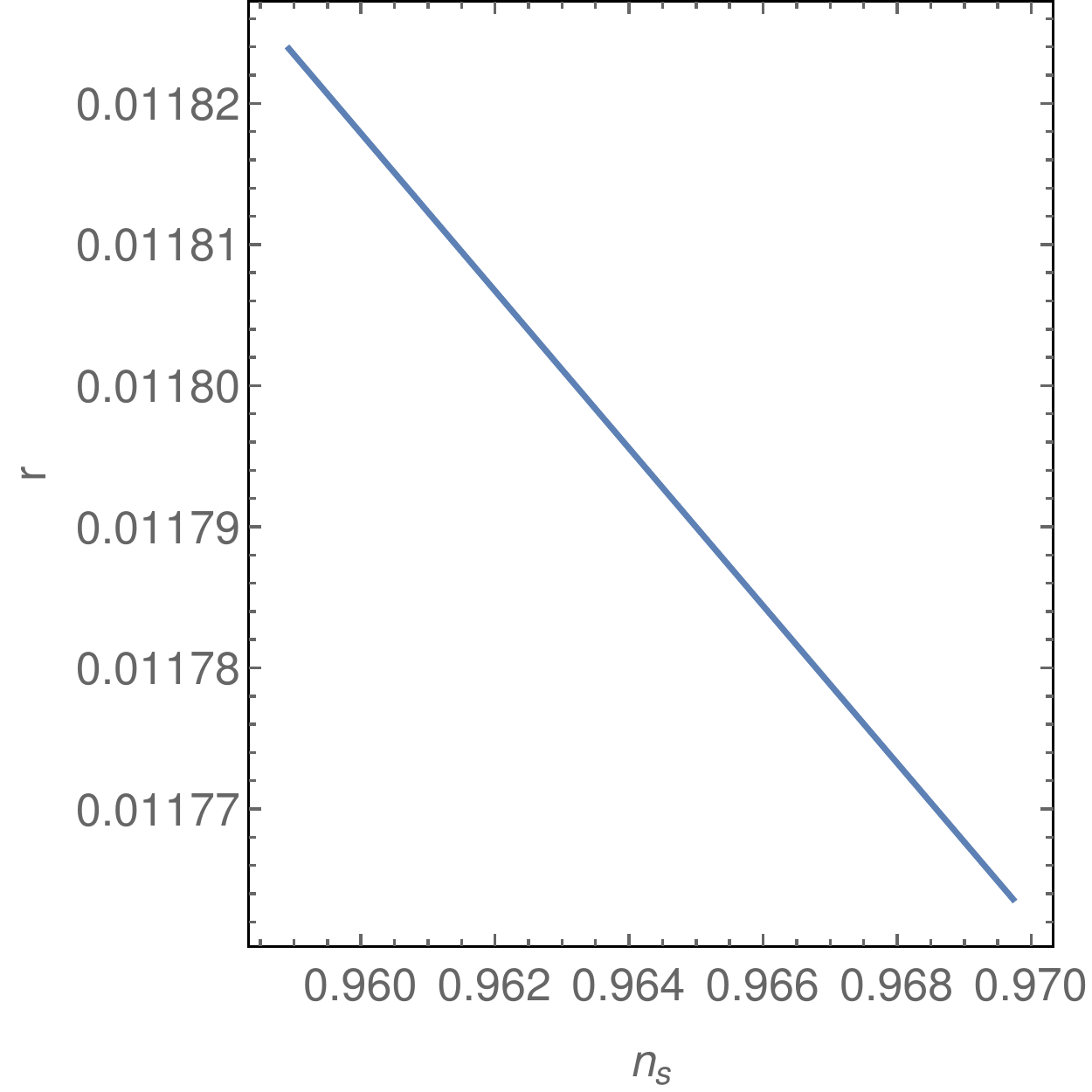}
\caption{Parametric plot of $n_s$ (along $x$-axis) vs. $r$ (along $y$-axis) 
with respect to $n$. Here we take $f_0 = 1$, $a_0 = 4$ and $n = [0.3062,0.3065]$.}
\label{plot-observable}
\end{center}
\end{figure}

It turns out that the theoretical predictions of scalar spectral index and the tensor-to-scalar ratio in the present case 
get simultaneously compatible with the Planck 2018 data for 
a small range of the parameters, given by: $f_0 = 1$, $a_0 = 4$ and $n = [0.3062,0.3065]$. Therefore the viable range of $n$ seems to be less than 
that of in the matter bounce scenario where $n = 1/3$; this result is in agreement with \cite{Elizalde:2020zcb}. The parametric plot $n_s$ vs. $r$ 
is depicted in the Fig.[\ref{plot-observable}].

Therefore in the context of ghost free Gauss-Bonnet theory of gravity -- the bouncing scenario in presence of a Type-IV singularity, 
where the Type-IV singularity $locally$ affects the spacetime around the time when the singularity occurs, 
turns out to be viable in respect to the Planck observations.

%%%%%%%%%%%%%%%%%%%%%%%%%%%%%%%%%%%%%%%%%%%%%%%%%%%%%%%%%%%%%%%%%%%%%%%%%%%%%%%%%%%%%%%%%%%%%%%%%%%%%%%%%%%%%%%%%%%%%%%%%%%%%%%%%%%%%%%%%%%%%88

\section{Conclusions}\label{sec_conclusion}

We have examined how the presence of a Type-IV singularity can influence the dynamics of a bouncing universe, namely, 
we have worked with bounce cosmology that appears with a Type-IV singularity at a finite time. 
In the case of a Type-IV singularity, the scale factor, the 
Hubble parameter and its first derivative are finite, however the higher derivatives of the Hubble parameter diverge at the time of of the singularity. 
Therefore the Type-IV singularity is not a crushing type, like the initial or the Big Rip singularity, and the universe can smoothly pass 
through a Type-IV singularity. However the presence of such a Type-IV singularity can severely influence the dynamics of the universe, as we have 
found here for an otherwise non-singular bounce scenario. The theory of gravity has been 
considered to be the well known ghost free Gauss-Bonnet (GB) gravity, 
where the ghost free nature is ensured by the presence of a Lagrange multiplier in the gravitational action, as developed in \cite{Nojiri:2018ouv}. 
Moreover we have chosen 
a class of Gauss-Bonnet coupling function ($h(t)$) that satisfies a constraint equation like 
$\ddot{h} = \dot{h}H$ (where $H$ is the Hubble parameter of the universe), which in turn leads to 
the speed of the gravitational wave as unity, and consequently, 
the model becomes compatible with the event GW170817. By using the reconstruction technique, we obtain
the explicit form of scalar field potential as well as the GB coupling function, which triggers a bouncing scenario with a Type-IV singularity 
at a finite time $t=t_s$.\\ 

We have found three different cases depending on whether $t_s<0$ or $t_s>0$ or $t_s = 0$ respectively -- 
(1) for $t_s < 0$, the bounce seems to happen at some negative time ($t_b < 0$, where the suffix stands for bounce) 
and the singularity occurs before the bounce, (2) for $t_s > 0$, the bounce 
shows at a positive time ($t_b>0$), and moreover, the singularity occurs after the bounce, and finally, (3) for $t_s = 0$, the bounce 
and the Type-IV singularity occur at the same instant of time, i.e $t_b = t_s = 0$. This is the first indication that the Type-IV singularity 
indeed affects the dynamics of the bouncing scenario. Consequently, we have analyzed the evolutions of the scalar and tensor perturbations in this context, 
and have determined various observable quantities like the scalar spectral index ($n_s$) and the tensor-to-scalar ratio ($r$) -- both of them are found to depend 
on $t_s$. Therefore the perturbation power spectra become different depending on the cases whether $t_s < 0$ or $t_s > 0$ or 
$t_s = 0$ respectively. However in all of these cases, the 
scalar power spectrum turns out to be highly red tilted and the tensor-to-scalar ratio becomes too large to be 
consistent with the Planck data. By a rigorous analysis of the scalar and the tensor perturbation, we have showed that it is difficult to 
obtain $n_s$ and $r$ matching upto the observed CMB spectra, in the case when the Type-IV singularity 
globally affects the spacetime. 
Such inconsistency of the observable quantities arises due to the appearance of the Type-IV singularity which ``globally'' 
affects the evolution of the Hubble parameter compared to the bouncing scenario where such a Type-IV singularity is absent. By the term ``global'', we 
mean that although the singularity occurs at a finite time $t = t_s$, it significantly affects the asymptotic evolution of the universe at 
the distant past as well as at the distant future. In particular, 
the presence of the Type-IV singularity results to the comoving Hubble radius going to zero asymptotically at both sides of the bounce, in effect of which, 
the perturbation modes generate near the bounce when all the relevant modes lie within the sub-Hubble regime. This is the reason 
why the scalar power spectrum shows a red tilted behaviour and the tensor-to-scalar ratio becomes too large in respect to the observational constraints.\\

Based on the above findings, we have investigated a different bouncing scenario which appears with a Type-IV singularity, however the Type-IV 
singularity ``locally'' affects the spacetime around the time when the singularity occurs. As a result, and unlike to the previous scenario, 
the comoving Hubble radius in this bounce scenario diverges to infinity at the distant past and thus the primordial perturbation modes generate 
far away from the bounce in the deep contracting phase. We have calculated the scalar spectral index and the tensor-to-scalar in this context, 
which are found to be simultaneously compatible with the recent Planck data for suitable regime of parameter values. This 
ensures the viability of the bounce model where the Type-IV singularity shows local effects on the spacetime around the time of the singularity. 
Here it is important to mention that the scalar tilt as well as the tensor-to-scalar ratio are found to be independent of the parameter $t_s$. 
This is however expected, because the singularity affects the spacetime locally around the time $t = t_s$ 
and the perturbation modes generate in the deep contracting phase where the singularity 
provides almost no effects on the universe's evolution.\\

Thus as a whole, this work clearly reveals that the presence of a Type-IV singularity has significant effects on an otherwise non-singular bounce scenario. 
We have showed that the bounce model that appears with a Type-IV singularity is viable if the singularity ``locally'' affects the spacetime 
around the time when it occurs, otherwise the observable quantities are found to be problematic (with respect to the Planck constraints) when the Type-IV 
singularity ``globally'' affects the spacetime. Therefore, in the realm of bouncing cosmology, if the universe faced a Type-IV 
singularity in the past during its evolution, then the singularity should ``locally'' affect the spacetime.

\section*{Acknowledgments}

This work was supported in part by MINECO (Spain), project PID2019-104397GB-I00 (SDO). 
This work was partially supported by the program Unidad de Excelencia Maria 
de Maeztu CEX2020-001058-M. This research was also supported in part by the 
International Centre for Theoretical Sciences (ICTS) for the online program - Physics of the Early Universe (code: ICTS/peu2022/1) (TP).

\end{document}